# An Investigation on the Basic Conceptual Foundations of Quantum Mechanics by Using the Clifford Algebra

**Elio Conte**

School of Advanced International Studies for Applied Theoretical and Non Linear Methodologies of Physics, Bari, Italy
Department of Human Physiology – TIRES – Center for Innovative Technologies for Signal Detection and Processing, Department of Physics-University of Bari- Italy.
elio.conte@fastwebnet.it

**Abstract:**
We review our approach to quantum mechanics adding also some new interesting results. We start by giving proof of two important theorems on the existence of the $A(Si)$ and $N_{i,\pm 1}$ Clifford algebras. This last algebra gives proof of the von Neumann basic postulates on the quantum measurement explaining thus in an algebraic manner the wave function collapse postulated in standard quantum theory. In this manner we reach the objective to expose a self-consistent version of quantum mechanics. In detail we realize a bare bone skeleton of quantum mechanics recovering all the basic foundations of this theory on an algebraic framework. We give proof of the quantum like Heisenberg uncertainty relations using only the basic support of the Clifford algebra. In addition we demonstrate the well known phenomenon of quantum Mach Zender interference using the same algebraic framework, as well as we give algebraic proof of quantum collapse in some cases of physical interest by direct application of the theorem that we derive to elaborate the $N_{i,\pm 1}$ algebra. We also discuss the problem of time evolution of quantum systems as well as the changes in space location, in momentum and the linked invariance principles. We are also able to re-derive the basic wave function of standard quantum mechanics by using only the Clifford algebraic approach. In this manner we obtain a full exposition of standard quantum mechanics using only the basic axioms of Clifford algebra.
we also discuss more advanced features of quantum mechanics . In detail, we give demonstration of the Kocken-Specher theorem, and also we give an algebraic formulation and explanation of the EPR paradox only using the Clifford algebra. By using the same approach we also derive Bell inequalities. Our formulation is strongly based on the use of idempotents that are contained in Clifford algebra. Their counterpart in quantum mechanics is represented by the projection operators that, as it is well known, are interpreted as logical statements, following the basic von Neumann results. Von Neumann realized a matrix logic on the basis of quantum mechanics. Using the Clifford algebra we are able to invert such result. According to the results previously obtained by Orlov in 1994, we are able to give proof that quantum mechanics derives from logic. We show that indeterminism and quantum interference have their origin in the logic. Therefore, it seems that we may conclude that quantum mechanics, as it appears when investigated by the Clifford algebra, is a two-faced theory in the sense that it looks from one side to "matter per se", thus to objects but simultaneously also to conceptual entities.
We advance the basic conclusion of the paper: ***There are stages of our reality in which we no more can separate the logic ( and thus cognition and thus conceptual entity) from the features of "matter per se". In quantum mechanics the logic, and thus the cognition and thus the conceptual entity-cognitive performance, assume the same importance as the features of what is being described. We are at levels of reality in which the truths of logical statements about dynamic variables become dynamic variables themselves so that a profound link is established from its starting in this theory between physics and conceptual entities***.
***Finally, in this approach there is not an absolute definition of logical truths. Transformations , and thus … "redefinitions"…. of truth values are permitted in such scheme as well as the well established invariance principles, clearly indicate***

## 1. A bare bone skeleton of quantum mechanics realized by Clifford algebra.

Let us start with a proper definition of the 3-D space Clifford (geometric) algebra $Cl_3$ .
It is an associative algebra generated by three vectors $e_1, e_2,$ and $e_3$ that satisfy the orthonormality relation

$$e_j e_k + e_k e_j = 2\delta_{jk} \qquad \text{for } j,k,\lambda \in [1,2,3] \qquad (1.1)$$

That is,

$$e_\lambda^2 = 1 \quad \text{and} \quad e_j e_k = -e_k e_j \quad \text{for } j \neq k$$

Let **a** and **b** be two vectors spanned by the three unit spatial vectors in $Cl_{3,0}$. By the orthonormality relation the product of these two vectors is given by the well known identity: $ab = a \cdot b + i(a \times b)$ where $i = e_1 e_2 e_3$ is a Clifford algebraic representation of the imaginary unity that commutes with vectors.
The (1.1) are well known in quantum mechanics. Here we give proof under an algebraic profile. Let us follow the approach that, starting with 1981, was developed by Y. Ilamed and N. Salingaros [1].
Let us admit that the three abstract basic elements, $e_i$, with $i = 1,2,3$ admit the following two postulates:

a) it exists the scalar square for each basic element:

$$e_1 e_1 = k_1 \text{ , } e_2 e_2 = k_2 \text{, } e_3 e_3 = k_3 \text{ with } k_i \in \Re \text{ .} \qquad (1.2)$$

In particular we have also the unit element, $e_0$, such that that

$e_0 e_0 = 1$, and $e_0 e_i = e_i e_0$

b) The basic elements $e_i$ are anticommuting elements, that is to say:

$e_1e_2 = -e_2e_1$ , $e_2e_3 = -e_3e_2$ , $e_3e_1 = -e_1e_3$ . (1.3)

**Theorem n.1.**

**Assuming the two postulates given in (a) and (b) with $k_i = 1$, the following commutation relations hold for such algebra :**

$e_1e_2 = -e_2e_1 = ie_3$ ; $e_2e_3 = -e_3e_2 = ie_1$; $e_3e_1 = -e_1e_3 = ie_2$ ; $i = e_1e_2e_3$,

$(e_1^2 = e_2^2 = e_3^2 = 1)$ (1.4)

**They characterize the Clifford Si algebra. We will call it the algebra A(Si).**

Proof.

Consider the general multiplication of the three basic elements $e_1, e_2, e_3$, using scalar coefficients $\omega_k, \lambda_k, \gamma_k$ pertaining to some field:

$e_1e_2 = \omega_1 e_1 + \omega_2 e_2 + \omega_3 e_3$ ; $e_2e_3 = \lambda_1 e_1 + \lambda_2 e_2 + \lambda_3 e_3$ ;

$e_3e_1 = \gamma_1 e_1 + \gamma_2 e_2 + \gamma_3 e_3$ . (1.4a)

Let us introduce left and right alternation: for any $(i,j)$, associativity exists $e_ie_ie_j = (e_ie_i)e_j$ and $e_ie_je_j = e_i(e_je_j)$ that is to say

$e_1e_1e_2 = (e_1e_1)e_2$ ; $e_1e_2e_2 = e_1(e_2e_2)$ ; $e_2e_2e_3 = (e_2e_2)e_3$ ; $e_2e_3e_3 = e_2(e_3e_3)$ ;

$e_3e_3e_1 = (e_3e_3)e_1$ ; $e_3e_1e_1 = e_3(e_1e_1)$ . (1.5)

Using the (1.4) in the (1.5) it is obtained that

$k_1e_2 = \omega_1 k_1 + \omega_2 e_1e_2 + \omega_3 e_1e_3$ ; $k_2e_1 = \omega_1 e_1e_2 + \omega_2 k_2 + \omega_3 e_3e_2$ ;

$k_2e_3 = \lambda_1 e_2e_1 + \lambda_2 k_2 + \lambda_3 e_2e_3$ ; $k_3e_2 = \lambda_1 e_1e_3 + \lambda_2 e_2e_3 + \lambda_3 k_3$ ;

$k_3e_1 = \gamma_1 e_3e_1 + \gamma_2 e_3e_2 + \gamma_3 k_3$ ; $k_1e_3 = \gamma_1 k_1 + \gamma_2 e_2e_1 + \gamma_3 e_3e_1$ . (1.6)

From the (1.6), using the assumption (b), we obtain that

$$\frac{\omega_1}{k_2}e_1e_2 + \omega_2 - \frac{\omega_3}{k_2}e_2e_3 = \frac{\gamma_1}{k_3}e_3e_1 - \frac{\gamma_2}{k_3}e_2e_3 + \gamma_3 ;$$

$$\omega_1 + \frac{\omega_2}{k_1}e_1e_2 - \frac{\omega_3}{k_1}e_3e_1 = -\frac{\lambda_1}{k_3}e_3e_1 + \frac{\lambda_2}{k_3}e_2e_3 + \lambda_3 ;$$

$$\gamma_1 - \frac{\gamma_2}{k_1}e_1e_2 + \frac{\gamma_3}{k_1}e_3e_1 = -\frac{\lambda_1}{k_2}e_1e_2 + \lambda_2 + \frac{\lambda_3}{k_2}e_2e_3 \quad (1.7)$$

We have that it must be

$\omega_1 = \omega_2 = \lambda_2 = \lambda_3 = \gamma_1 = \gamma_3 = 0$ (1.8)

and

$-\lambda_1 k_1 + \gamma_2 k_2 = 0$ $\gamma_2 k_2 - \omega_3 k_3 = 0$ $\lambda_1 k_1 - \omega_3 k_3 = 0$ (1.9)

The following set of solutions is given:

$k_1 = -\gamma_2\omega_3$, $k_2 = -\lambda_1\omega_3$ , $k_3 = -\lambda_1\gamma_2$ (1.10)

that is to say

$\omega_3 = \lambda_1 = \gamma_2 = i$ (1.11)

In this manner , as a theorem, the existence of such algebra is proven. The basic features of this algebra are given in the following manner

$$e_1^2 = e_2^2 = e_3^2 = 1 \ ; \ e_1e_2 = -e_2e_1 = ie_3 \ ; \ e_2e_3 = -e_3e_2 = ie_1; \ e_3e_1 = -e_1e_3 = ie_2 \ ;$$

$$i = e_1e_2e_3 \tag{1.12}$$

The content of the theorem n.1 is thus established: given three abstract basic elements as defined in (a) and (b) ($k_i = 1$), an algebraic structure is given with four generators ($e_0, e_1, e_2, e_3$).

Note that in the algebra A (Si) the $e_i$ ($i = 1,2,3$) have an intrinsic potentiality that is to say an ontic potentiality or equivalently an irreducible intrinsic indetermination. Since $e_i^2 = 1$ ($i = 1,2,3$) , we may think to attribute them or the numerical value +1 or the numerical value –1 .

A generic member of our algebra A(Si) is given by

$$x = \sum_{i=0}^{4} x_i e_i = x_0 + \mathbf{x} \tag{1.13}$$

with $x_i$ pertaining to some field $\Re$ or $C$ .

We may define [2] the hyperconjugate $\hat{x}$

$$\hat{x} = x_0 - \mathbf{x}$$

the complex conjugate

$$x^* = x_0^* + \mathbf{x}^\circ$$

and the conjugate

$$\bar{x} = x_0^* - \mathbf{x}^*$$

The *Norm* of x is defined as

$$Norm\,(\mathrm{x}) = x\,\hat{x} = \hat{x}\,x = x_0^2 - x_1^2 - x_2^2 - x_3^2 \tag{1.14}$$

with

*Norm (xy) = Norm (x) Norm (y)*

The proper inverses of the basic elements $e_i$ ($i = 1,2,3$) are themselves. Given the member x , its inverse $x^{-1}$

is $\hat{x} / Norm$ (x) with *Norm* (x) $\neq 0$

We may transform Clifford members according to Linear Transformations

$$x' = AxB + C \tag{1.14a}$$

with unitary norms for the employed Clifford members $A, B$ and $C = 0$ for linear homogeneous transformation.

Let us now take a step on.

As previously said, in the algebra A (Si) the $e_i$ ($i = 1,2,3$) have an intrinsic potentiality that is to say an ontic potentiality or equivalently an irreducible intrinsic indetermination. Since $e_i^2 = 1$ ($i = 1,2,3$), we may think to attribute them or the numerical value +1 or the numerical value –1. Let us give proof of such our basic assumption.

Since the $e_i$ are abstract entities, having the potentiality that we may think to attribute them the numerical values, $\pm 1$, they have an intrinsic and irreducible indetermination. Therefore, we may admit to be $p_1(+1)$ the probability that $e_1$ assumes the value $(+1)$ and $p_1(-1)$ the probability that it assumes the value $-1$. We may represent the mean value that is given by

$$< e_1 >= (+1)p_1(+1) + (-1)p_1(-1) \tag{1.15}$$

Considering the same corresponding notation for the two remaining basic elements, we may introduce the other two mean values:

$$< e_2 >= (+1)p_2(+1) + (-1)p_2(-1), \tag{1.16}$$

$$< e_3 >= (+1)p_3(+1) + (-1)p_3(-1).$$

We have

$$-1 \le < e_i > \le +1 \quad i = (1,2,3) \tag{1.17}$$

Selected the following generic element of the algebra A(Si):

$$x = \sum_{i=1}^{3} x_i e_i \qquad x_i \in \Re \tag{1.18}$$

Note that

$$x^2 = x_1^2 + x_2^2 + x_3^2 \tag{1.19}$$

Its mean value results to be

$$< x >= x_1 < e_1 > + x_2 < e_2 > + x_3 < e_3 > \tag{1.20}$$

Let us call

$$b = (x_1^2 + x_2^2 + x_3^2)^{1/2} \tag{1.21}$$

so that we may attribute to $x$ the value $+b$ or $-b$

We have that

$$-b \le x_1 < e_1 > + x_2 < e_2 > + x_3 < e_3 > \le b \tag{1.22}$$

The (1.22) must hold for any real number $x_i$, and, in particular, for

$$x_i =< e_i >$$

so that we have that

$$x_1^2 + x_2^2 + x_3^2 \le b$$

that is to say

$$b^2 \le b \quad \rightarrow b \le 1$$

so that we have the fundamental relation

$$< e_1 >^2 + < e_2 >^2 + < e_3 >^2 \le 1 \tag{1.23}$$

This is a basic relation of irreducible indetermination that we are writing in our Clifford algebraic elaboration. Of course it is well known in quantum theory [3]. Let us observe some important features:

(a) In absence of numerical attribution to the $e_i$ ( and in analogy with physics this means ….in absence of a measurement, that is to say in absence of direct observation of one quantum observable), the (1.23) holds. If we attribute instead a definite numerical value to one of the three entities, as example we attribute to $e_3$ the numerical value $+1$ , we have

$< e_3 >= 1$, and the (1.23)) is reduced to

$$< e_1 >^2 + < e_2 >^2 = 0 , < e_1 >=< e_2 >= 0 , \qquad (1.24)$$

and we have complete, irreducible, indetermination for $e_1$ and for $e_2$ .

( b ) Finally, the (1.23) affirms that we never can attribute simultaneously definite numerical values to two basic non commutative elements $e_i$ .

We may now summarize the obtained results. First, we retain that the first axiom of the $A(Si)$ algebra, the (1.2) with

$k_i = 1$, indicates that the abstract basic elements $e_i$ have an ontic potentiality, that is to say that they have an irreducible indeterminism as supported finally from the (1.23). In order to characterize such features we have introduced the concept of mean value for such algebraic entities and, consequently, that one of potentiality. When we attempt to attribute a numerical values to an abstract element, as it happens as example, in the (1.24), we perform an operation that in physics has a counterpart that is called an act of measurement. For us, any measurement is a semantic act. No matter if the measurement is performed by a technical instrument or by an human observer. In any case it is realized having at its basic arrangement, a semantic act. If we remain in the restricted domain of the $A(Si)$, we are in some sense in a condition that, on the general plane, may be assimilated to that one in which we have human or technical systems that are in some manner forced to answer to questions (the attribution of numerical values to the basic elements) which they cannot understand in line of principle. As consequence, the probabilities that we have used in the (1.15) and in the (1.16) are fundamentally different from classical probabilities under a basic conceptual profile. In classical probability theory, as it is known, probabilities represent a lack of information about preexisting and pre-established properties of systems .In the present case we have instead a situation in which we have not an algorithm in $A(Si)$ to execute a semantic act devoted to identify the meaning of a statement in terms of truth values and in relation to another statement. So we need to introduce probabilities.

## 1.1 Derivation of a quantum like Heisenberg Uncertainty Relation by Using Only Clifford Algebra.

T.F. Jordan published a book on Quantum mechanics in simple matrix form [3] in which this author derived previously Heisenberg's uncertainty relation using matrices.

Consider here the following two members of Clifford algebra

$$k \equiv k(e_1, e_2, e_3) = \alpha e_1 + \beta e_2 + \gamma e_3 ;$$

$$L \equiv L(e_1, e_2, e_3) = a e_1 + b e_2 + c e_3 ; \qquad (1.25)$$

with

$(\alpha, \beta, \gamma) \in \Re$ and $(a,b,c) \in \Re$

We have that

$$k^2 = \alpha^2 + \beta^2 + \gamma^2 \geq 0 \qquad \text{and} \quad L^2 = a^2 + b^2 + c^2 \geq 0 \tag{1.26}$$

$k$ and $L$ may assume respectively the numerical values

$$\pm\sqrt{\alpha^2 + \beta^2 + \gamma^2} \quad \text{and} \quad \pm\sqrt{a^2 + b^2 + c^2} \tag{1.27}$$

Let us consider the following Clifford member

$$U = (k + wL)(K + w^*L) = k^2 + ww^*L^2 + a(Lk + kL) + ib(Lk - kL) \tag{1.28}$$

where

$(w = q + ip) \in C$

It is

$$kL + Lk = 2\alpha a + 2\beta b + 2\gamma c \tag{1.29}$$

and

$$Lk - kL = (2b\gamma - 2c\beta)ie_1 + (2c\alpha - 2a\gamma)ie_2 + (2a\beta - 2b\alpha)ie_3 \tag{1.30}$$

Consider

$q = 0$ and $p = 1$

We have the Clifford algebraic element

$$U - k^2 - L^2 = Se_1 + Re_2 + Te_3 \tag{1.31}$$

where

$$S = 2c\beta - 2b\gamma \; ; \quad R = 2a\gamma - 2c\alpha \quad ; \quad T = 2b\alpha - 2a\beta \tag{1.32}$$

The mean value of such algebraic element is

$$\langle U \rangle = \langle k^2 \rangle + \langle L^2 \rangle + \langle Se_1 + Re_2 + Te_3 \rangle \tag{1.33}$$

Let us consider three real numbers $(m,n,r)$ such that

$$x + y = -\frac{m}{n} \quad \text{and} \quad xy = \frac{r}{n} \quad \text{with} \quad x = \langle k^2 \rangle \quad \text{and} \quad y = \langle L^2 \rangle$$

we have that

$x + y = sxy$ with $s = -m/r$

Consequently, we have that

$$\langle U \rangle = s\langle k^2 \rangle \langle L^2 \rangle + \langle Se_1 + Re_2 + Te_3 \rangle \tag{1.34}$$

Note that it is

$$\langle k^2 \rangle \langle L^2 \rangle = (\alpha^2 + \beta^2 + \gamma^2)(a^2 + b^2 + c^2) = \omega + \lambda \tag{1.35}$$

where

$$\omega = b^2\alpha^2 + c^2\alpha^2 + a^2\beta^2 + c^2\beta^2 + a^2\gamma^2 + b^2\gamma^2 \tag{1.36}$$

and

$$\lambda = a^2\alpha^2 + b^2\beta^2 + c^2\gamma^2 \tag{1.37}$$

Of course , according to our Clifford algebra, we have that

$$(Se_1 + Re_2 + Te_3)\,(Se_1 + Re_2 + Te_3) = S^2 + R^2 + T^2 \tag{1.38}$$

and this is to say that

$$-\sqrt{S^2+R^2+T^2} \le < Se_1 + Re_2 + Te_3 > \le \sqrt{S^2+R^2+T^2} \quad (1.39)$$

In conclusion, the mean value $< Se_1 + Re_2 + Te_3 >$ is a number enclosed between

$$\pm 2\sqrt{\omega - 2bc\beta\gamma - 2ac\alpha\gamma - 2ab\alpha\beta} . \quad (1.40)$$

Therefore, always we have that

$$< k^2 >< L^2 > \ \ge \ \left| \frac{1}{2} < kL - Lk > \right|^2 . \quad (1.41)$$

Consider now the two following Clifford members

$$A = k + < A > \quad \text{and} \quad B = L + < B > \quad (1.42)$$

where $k$ and $L$ have been given in (1.25).

It is

$$kL - Lk = AB - BA \quad (1.43)$$

Inserting the (1.40)) and the (1.43) in the (1.41), we obtain that

$$< (A - < A >)^2 >< (B - < B >)^2 > \ge \left| \frac{1}{2} < AB - BA > \right|^2 \quad (1.44)$$

This is the standard Clifford algebraic expression of Heisenberg uncertainty quantum like relation in quantum mechanics.

Note the salient feature that it has obtained only by algebraic means without recovering any concrete physics. In particular we do not recall here a quantization of canonically conjugate variables and Planck constant, $\hbar$ .

## 1.2 Proof that Quantum Interference Arises in a Clifford Algebraic Formulation of a Bare Bone Skeleton of Quantum Mechanics and the irreducible, ontic randomness of the basic Clifford algebraic elements.

This time let us take a step on and consider a pressing physical argument.

Consider a beam of particles impinging on a beam splitter $A$ so that randomly may be either reflected to proceed a path $L_1$ or transmitted to proceed along the path $L_2$ (Fig.1). At the end of $L_1$, the particles impinge on the upper side of a second beam splitter, $B$, and it may be either reflected and detected by the detector $D_1$ or transmitted and detected by the detector $D_2$ .The particles arriving from path $L_2$, impinge on the opposite side to be either transmitted reaching the detector $D_1$ or reflected to reach the counter $D_2$ .As it is well known we are considering here the interference pattern of a beam of particles passing through a Mach Zender interferometer.

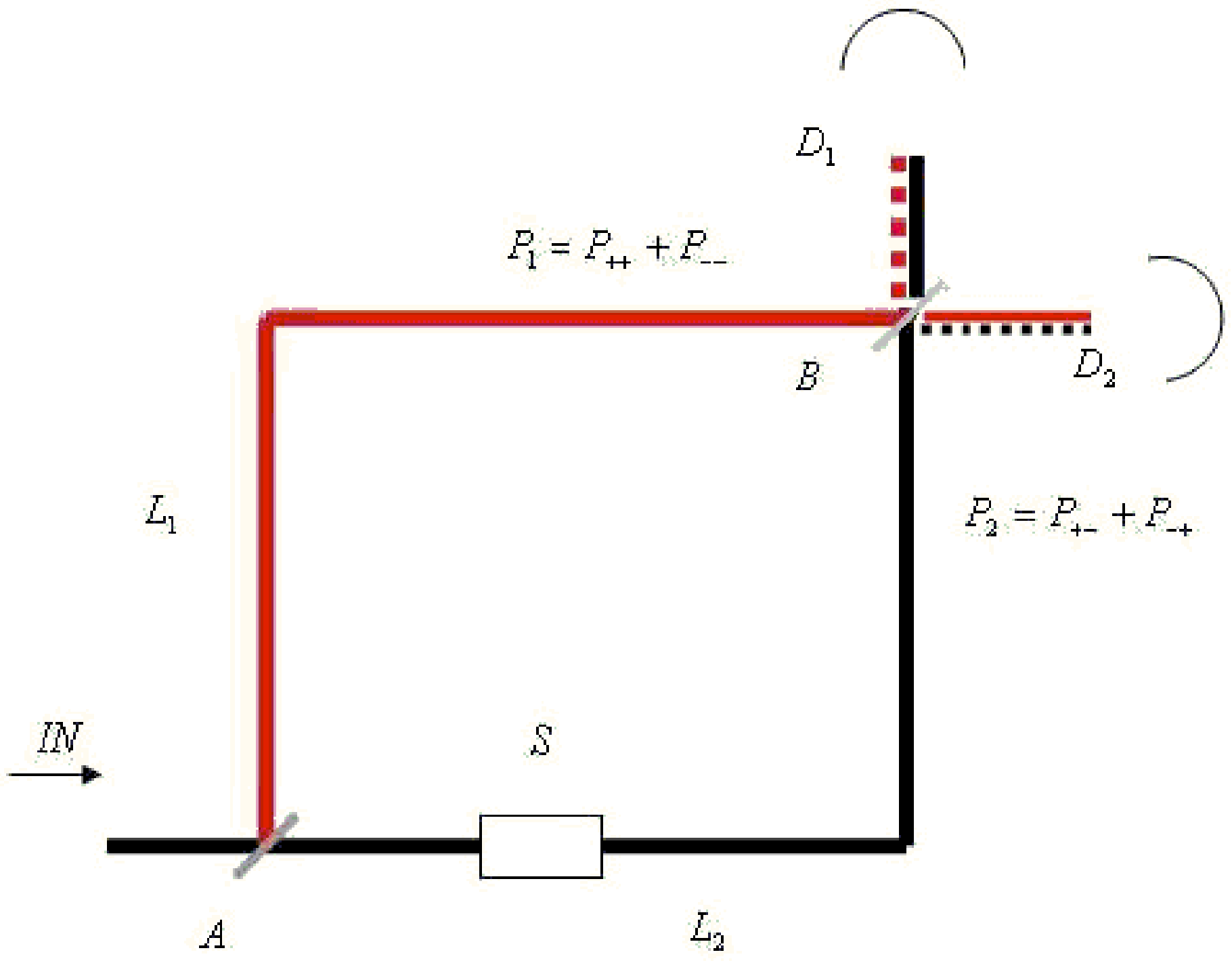

Of course this is a standard argument in quantum mechanics and also so long debated. Rather recently it has been exposed in the present form in a previous paper [4].

Let us modify lightly the basic framework of the experiment in the sense that, instead of considering its basic physical features , let us consider it from a logical point of view and thus under the profile of an human cognitive-conceptual performance. Consider a logic entity represented by the random variable $A$. It may assume the value $a = +1$ in the case of reflection and the value $a = -1$ in the case of transmission. The same argument holds for the random variable $B$. It assumes the value $b = +1$ in the case of reflection and the value $b = -1$ in the case of transmission. We have a third variable $C = AB$ that is determined by the product of the values of $A$ and B.

In analogy with the rough quantum scheme previously developed , we call still write the mean value of $A$ by $< A >$ and

$$< A >= (a = +1) p_{ab} + (a = -1) p_{ab} \tag{1.45}$$

the mean value of $B$ by $< B >$ and

$$< B >= (b = +1) p_{ab} + (b = -1) p_{ab} \tag{1.46}$$

and the mean value of $C$ by $< C >$ and

$$\begin{aligned} < C >= (ab; a = +1, b = +1) p_{ab} + (ab; a = +1, b = -1) p_{ab} + \\ (ab; a = -1, b = +1) p_{ab} + (ab; a = -1, b = -1) p_{ab} \end{aligned} \tag{1.47}$$

Now let us expose the argument as it was recently developed in [4,5]. We may write easily the expression of the probability for the corresponding four logic alternatives ($a = \pm 1, b = \pm 1$) in the following manner

$$p_{ab} = \frac{1}{4}(1 + ax + by + abz) \qquad (1.47a)$$

where

$$x \equiv \langle A \rangle, \; y \equiv \langle B \rangle, z \equiv \langle C \rangle. \qquad (1.48)$$

For the probability for having cognition on counting the detector $D_1$, we have that

$$p_{++} = \frac{1}{4}(1 + x + y + z) \quad \text{and} \quad p_{--} = \frac{1}{4}(1 - x - y + z) \qquad (1.49)$$

so that in the detector $D_1$ we have

$$p_{D_1} = p_{++} + p_{--} = \frac{1}{2}(1 + z) = \frac{1}{2}(1 + \langle C \rangle) \qquad (1.50)$$

In the case of the detector $D_2$, we have

$$p_{+-} = \frac{1}{4}(1 + x - y - z) \quad \text{and} \quad p_{-+} = \frac{1}{4}(1 - x + y - z) \qquad (1.51)$$

and

$$p_{D_2} = p_{+-} + p_{-+} = \frac{1}{2}(1 - z) = \frac{1}{2}(1 - \langle C \rangle) \qquad (1.52)$$

This is of course the classical statistical argument holding on an epistemic interpretation of randomness . In order to introduce the quantum like elaboration (see also [4] ) , let us introduce three new variables:

$$U = \alpha A + \beta B + \gamma C \quad \text{with} \quad \alpha^2 + \beta^2 + \gamma^2 = 1; \qquad (1.52a)$$

$$V = \lambda A + \mu B + \nu C , \quad \text{with } \lambda^2 + \mu^2 + \nu^2 = 1 , \quad \alpha\lambda + \beta\mu + \gamma\nu = 0 \qquad (1.53)$$

and

$$W = \delta A + \omega B + \vartheta C \quad \text{with} \quad \delta = \beta\nu - \gamma\mu \; ; \; \omega = \gamma\lambda - \alpha\nu , \vartheta = \alpha\mu - \beta\lambda , \qquad (1.54)$$

and consider

$$\langle U \rangle = u \qquad (1.55)$$

plus

$$\langle V \rangle = \langle W \rangle = 0 \qquad (1.56)$$

in order to take into account a complete indetermination in the case of variables $V$ and $W$.

Following this development, one obtains

$$\alpha \langle A \rangle + \beta \langle B \rangle + \gamma \langle C \rangle = u \; ; \qquad (1.57)$$

$$\lambda \langle A \rangle + \mu \langle B \rangle + \nu \langle C \rangle = 0 ;$$

$$\delta \langle A \rangle + \omega \langle B \rangle + \vartheta \langle C \rangle = 0$$

that admit solutions

$< A >= \alpha u$ , $< B >= \beta u$ , $< C >= \gamma u$ . (1.58)

Inserting the (1.58) in the (1.47a), one obtains

$$p_{ab} = \frac{1}{4}\left[1 + (\alpha a + \beta b + \gamma ab)u\right] \quad (1.59)$$

and

$$p_{D_1} = p_{++} + p_{--} = \frac{1}{2}(1 + \gamma u) \quad (1.60)$$

and

$$p_{D_2} = p_{+-} + p_{-+} = \frac{1}{2}(1 - \gamma u) \quad (1.61)$$

Let us comment the obtained results.

First consider the classical case.

Probability given in (1.47a) must be between the well known limits

$$0 \le p_{ab} \le 1 \quad (1.62)$$

Consequently, still according to ref.[4], $< A >, < B >, < C >$, are the coordinates of a point inside the equilateral octahedron having the vertices

$< A >= \pm 1$, $< B >= 0$, $< C >= 0$; $< B >= \pm 1$, $< A >= 0$, $< C >= 0$; $< C >= \pm 1$, $< A >= 0$, $< C >= 0$

Physically speaking: The first limiting values correspond to the case of pure reflection (transmission) by $A$ and equally probable reflection and transmission by $B$ and zero correlation. The second limiting values correspond to equally probable reflection and transmission by $A$ followed by pure reflection by $B$ and zero correlation, and the third limiting values correspond to the case of complete correlation between the two splitters with equally probable transmission and reflection by $A$ and $B$. These are the limiting cases while for the other possible conditions we have average values of the considered random variables having values less than one. This means that always particles with both the values ($\pm 1$) are present. We have that

$$-1 \le < A > + < B > + < C > \le +1 \quad (1.63)$$

According to the (1.58) in the case of the (1.52)-(1.54) and (1.55)-(1.56), we have that

$$-1 \le (\alpha + \beta + \gamma)u \le +1 \quad (1.64)$$

which implies that the absolute value of $u$ is always smaller than one. Particles with relation to both logical values ($\pm 1$) of $A, B, C$ are always present.

We may now explore the quantum-like case. Instead of the (1.50) and of the (1.52), of the (1.60) and the (1.61), the correct probabilities in quantum theory result to be

$$p_{D_1} = p_{++} + p_{--} = \frac{1}{2}(1 + \gamma) \quad (1.65)$$

and

$$p_{D_2} = p_{+-} + p_{-+} = \frac{1}{2}(1-\gamma) \tag{1.66}$$

that result to be

$$p_{D_1} = p_{++} + p_{--} = \frac{1}{2}(1+\cos\phi) \tag{1.67}$$

and

$$p_{D_2} = p_{+-} + p_{-+} = \frac{1}{2}(1-\cos\phi) \tag{1.68}$$

This is to say that we must have $u = 1\,(u = -1)$, and

$$<A> = \alpha\,,\ <B> = \beta\,,\ <C> = \gamma \tag{1.69}$$

with

$$<A>^2 + <B>^2 + <C>^2 = 1 \tag{1.70}$$

and

$$\gamma = \cos\phi$$

as polar angle of the unit vector on the sphere given in (1.70). This is to say that it must be

$$<U^2> = 1 \tag{1.71}$$

and

$$<V> = <W> = 0 \tag{1.72}$$

to assure complete indetermination.

Let us consider again the variable $U$ as given in the (1.52a). It results that

$$U^2 = (\alpha A + \beta B + \gamma C)(\alpha A + \beta B + \gamma C) = \alpha^2 + \beta^2 + \gamma^2 + \alpha\beta(AB+BA) + \tag{1.73}$$

$$\alpha\gamma(AC+CA) + \beta\gamma(BC+CB) = 1 + \alpha\beta(AB+BA) + \alpha\gamma(AC+CA) + \beta\gamma(BC+CB)$$

It is

$$<U^2> = 1 + \ <\alpha\beta(AB+BA) + \alpha\gamma(AC+CA) + \beta\gamma(BC+CB)>. \tag{1.74}$$

The only way to obtain the (1.71) is that

$$AB = -BA, \qquad AC = -CA, \qquad BC = -CB \tag{1.75}$$

and this leads to only one possible conclusion :

it is to say that the variables $A, B, C$ must be the basic elements of the Clifford algebra, the $e_i$ $(i = 1,2,3)$ basic elements that we introduced above.

It is

$A \equiv e_1$ , $B \equiv e_2$, $AB \equiv e_1 e_2 = ie_3$

Therefore, $A, B, C$ , must be members of the Clifford algebra.

So we reach the following two conclusions:

a) Clifford algebraic structure is able to evidence the bare bone skeleton of quantum mechanics. Moreover, it is absolutely necessary to explain the basic foundations of quantum theory.
b) In 1932 J. von Neumann [6] showed that projection operators and, in particular, quantum density matrices can be interpreted as logical statements If $|\psi\rangle$ is a state vector for a quantum state in which the observable $S$ is equal to $k$, than the density matrix $|\psi\rangle\langle\psi|$ represents the logical statement $\Lambda_k$ which says "$S = k$". He demonstrated the possibility of constructing a matrix logic from quantum mechanics. According also to the basic work of Orlov, we have inverted here the situation. We have shown that quantum mechanics derives from logic. Remember that the main quantum phenomena are indeterminism and interference. We have shown here that both such basic features may be obtained on a purely algebraic Clifford construction and on a purely logical basis. We may reconsider the original Orlov statement "every atomic proposition of classic logic can be represented by a diagonal operator", the third Clifford basic element $e_3$ [7]. We will consider in detail such argument in the following sections.

Quantum mechanics holds about the basic phenomenon of quantum interference. We may realize it using the basic elements, and the structure of the Clifford algebra.

## 2. A Proof of von Neumann's postulates on quantum measurement by using the Clifford Algebra

Let us evidence another important feature of Clifford algebra $A(Si)$.

In Clifford algebra $A(Si)$ we have idempotents (this is to say…as counterpart we have projection operators in quantum mechanics).In von Neumann language projection operators can be interpreted as logical statements.

Let us give some example of idempotents in Clifford algebra.

It is well known the central role of density matrix in traditional quantum mechanics . In the Clifford algebraic scheme, we have a corresponding algebraic member that is given in the following manner

$$\rho = a + be_1 + ce_2 + de_3 \qquad (2.1)$$

with

$$a = \frac{|c_1|^2}{2} + \frac{|c_2|^2}{2}, \; b = \frac{c_1^* c_2 + c_1 c_2^*}{2}, \; c = \frac{i(c_1 c_2^* - c_1^* c_2)}{2}, \quad d = \frac{|c_1|^2 - |c_2|^2}{2} \qquad (2.2)$$

where the $e_i$ are the basic elements in our algebraic Clifford scheme while in matrix notation, $e_1$, $e_2$, and $e_3$ in standard quantum mechanics are the well known Pauli matrices. The complex coefficients $c_i$ ($i = 1,2$) are the well known probability amplitudes for the considered quantum state

$$\psi = \begin{pmatrix} c_1 \\ c_2 \end{pmatrix} \quad \text{and} \quad |c_1|^2 + |c_2|^2 = 1 \qquad (2.3)$$

For a pure state in quantum mechanics it is $\rho^2 = \rho$. In our scheme a theorem may be demonstrated in Clifford algebra. It is that

$$\rho^2 = \rho \leftrightarrow \ a = \frac{1}{2} \ \text{ and } \ a^2 = b^2 + c^2 + d^2 \tag{2.4}$$

The details of this our theorem are given in [8]. We have also $Tr(\rho) = 2a = 1$[8]. In this manner we have the necessary and sufficient conditions for $\rho$ to represent a Clifford member whose counterpart in standard quantum mechanics represents a potential state or, equivalently, a superposition of states.

Let us consider still other two of such idempotents in $A(Si)$

$$\psi_1 = \frac{1+e_3}{2} \quad \text{and} \quad \psi_2 = \frac{1-e_3}{2} \tag{2.5}$$

It is easy to verify that $\psi_1^2 = \psi_1$ and $\psi_2^2 = \psi_2$.

Let us examine now the following algebraic relations:

$$e_3\psi_1 = \psi_1 e_3 = \psi_1 \tag{2.6}$$

$$e_3\psi_2 = \psi_2 e_3 = -\psi_2 \tag{2.7}$$

Similar relations hold in the case of $e_1$ or $e_2$.

Here is one central aspect of the present paper. By a pure semantic act, looking at the (2.6) and (2.7), we reach only a conclusion. With reference to the idempotent $\psi_1$, the algebra A(Si) (see the (2.6)), attributes to $e_3$ the numerical value of $+1$ while, with reference to the idempotent $\psi_2$, the algebra A(Si) attributes to $e_3$ (see the (2.7)), the numerical value of -1 .

The basic point is that at the basis we have a semantic act .

However, assuming the attribution $e_3 \rightarrow +1$, from the (1.4) we have that new commutation relations should hold in a new Clifford algebra , given in the following manner :

$$e_1^2 = e_2^2 = 1, \quad i^2 = -1; \qquad e_1e_2 = i, \qquad e_2e_1 = -i, \qquad e_2 i = -e_1,$$
$$ie_2 = e_1, \qquad e_1 i = e_2, \qquad ie_1 = -e_2 \tag{2.8}$$

with three new basic elements ($e_1, e_2, i$) instead of ($e_1, e_2, e_3$) .

We totally agree with the possible criticism that such our argument to express the (2.8) on the basis of a rough attribution to $e_3$ is in itself very rough, and, in any case, only pertaining, still again, a pure semantic operation, and actually, as an adventure, we are attempting to admit that in the case in which we attribute to $e_3$ the numerical value +1, a new algebraic structure should arise with new generators whose rules should be given in (2.8) instead of the (1.4). Therefore, the arising central problem is that we should be able to proof the real existence of such new algebraic structure with rules given in the (2.8). We repeat: in the case of the starting algebraic structure, the algebra A(Si), we showed by theorem n.1 that it exists with its proper rules:

$$e_1^2 = e_2^2 = e_3^2 = 1;$$
$$e_1e_2 = -e_2e_1 = ie_3 \ ; \ e_2e_3 = -e_3e_2 = ie_1; \ e_3e_1 = -e_1e_3 = ie_2 \ ; i = e_1e_2e_3 \tag{2.9}$$

In the present case in which we attribute to $e_3$ the numerical value +1, we should demonstrate that it exists a new algebra given in the following manner

$e_1^2 = e_2^2 = 1$; $i^2 = -1$;

$e_1e_2 = i$ , $e_2e_1 = -i$, $e_2i = -e_1, ie_2 = e_1, e_1i = e_2$, $ie_1 = -e_2$ (2.10)

So we arrive to give proof of the theorem n.2.

**Theorem n.2 .**

**Assuming the postulates given in (a) and (b) with $k_1 = 1$, $k_2 = 1, k_3 = -1$, the following commutation rules hold for such new algebra:**

$e_1^2 = e_2^2 = 1$; $i^2 = -1$;

$e_1e_2 = i$ , $e_2e_1 = -i$, $e_2i = -e_1, ie_2 = e_1, e_1i = e_2$, $ie_1 = -e_2$ (2.11)

**They characterize the Clifford Ni algebra. We will call it the algebra $N_{i,+1}$**

Proof

To give proof, rewrite the (1.4a) in our case, and performing step by step the same calculations of the previous proof, we arrive to the solutions of the corresponding homogeneous algebraic system that in this new case are given in the following manner:

$k_1 = -\gamma_2\omega_3$; $k_2 = -\lambda_1\omega_3; k_3 = -\lambda_1\gamma_2$ (2.12)

where this time it must be $k_1 = k_2 = +1$ and $k_3 = -1$. It results

$\lambda_1 = -1; \gamma_2 = -1$; $\omega_3 = +1$ (2.13)

and the proof is given.

The content of the theorem n.2 is thus established. **When we attribute to $e_3$ the numerical value +1 we pass from the Clifford algebra $A(Si)$ to a new Clifford algebra $N_{i,+1}$ whose algebraic structure is no more given from the (2.9) of the algebra $A(Si)$ but from the following new basic rules:**

$e_1^2 = e_2^2 = 1$; $i^2 = -1$;

$e_1e_2 = i$ , $e_2e_1 = -i$, $e_2i = -e_1, ie_2 = e_1, e_1i = e_2$, $ie_1 = -e_2$ (2.14)

The theorem n.2 also holds in the case in which we attribute to $e_3$ the numerical value of $-1$ .

**Assuming the postulates given in (a) and (b) with $k_1 = 1$, $k_2 = 1, k_3 = -1$, the following commutation rules hold for such new algebra**

$e_1^2 = e_2^2 = 1$; $i^2 = -1$;

$e_1e_2 = -i$ , $e_2e_1 = i$, $e_2i = e_1, ie_2 = -e_1, e_1i = -e_2$, $ie_1 = e_2$ (2.15)

**They characterize the Clifford Ni algebra. We will call it the algebra $N_{i,-1}$**

To give proof , consider the solutions of the (2.12) that are given in this new case by

$\lambda_1 = +1; \gamma_2 = +1$; $\omega_3 = -1$ (2.16)

and the proof is given.

The content of the theorem n.2 is thus established. When we attribute to $e_3$ the numerical value –1, we pass from the Clifford algebra $A(Si)$ to a new Clifford algebra $N_{i,-1}$ whose algebraic structure is not given from the (2.9) of the algebra $A(Si)$ and not even from the (2.14) but from the following new basic rules:

$$e_1^2 = e_2^2 = 1;\; i^2 = -1;$$

$$e_1 e_2 = -i \quad , \quad e_2 e_1 = i, \quad e_2 i = e_1, ie_2 = -e_1, e_1 i = -e_2, \quad ie_1 = e_2 \qquad (2.17)$$

In a similar way, proofs may be obtained when we consider the cases attributing numerical values ($\pm 1$) to $e_1$ or to $e_2$.

The Clifford algebra, $N_{1,\pm 1}$, given in the (2.15 ) and in the (2.17) are the dihedral Clifford algebra $N_i$ (for details, see ref.[1]).

In conclusion, we have shown two basic theorems, the theorem n.1 and the theorem n.2. As any mathematical theorem they have maximum rigour, and an aseptic mathematical content that cannot be questioned. The basic statement that we reach by the proof of such two theorems is that in Clifford algebraic framework, we have the Clifford algebra A(Si) and inter-related Clifford algebras $N_{i,\pm 1}$. When we consider $(e_1, e_2, e_3)$ as the three abstract elements with rules given in (2.9) , we are in the Clifford algebra A(Si) .When we attribute to $e_3$ the numerical value +1, we pass from the algebra $A(Si)$ to the Clifford algebra $N_{i,+1}$. Instead, when we pass from the Clifford algebra $A(Si)$ to the Clifford algebra $N_{i,-1}$, we are attributing to $e_3$ the numerical value –1 .

The same conceptual facts hold when we reason for Clifford basic elements $e_1$ or to $e_2$, attributing in this case a possible numerical value ($\pm 1$) or to $e_1$ or to $e_2$, respectively.

Obviously the implications of such shown theorems for the measurement problem in quantum mechanics are of relevant interest.

If one looks at the algebraic rules and commutation relations given in the (2.9), the algebra $A(Si)$ immediately remembers that they are universally valid in quantum mechanics. It links the Pauli matrices that are sovereign in quantum theory. Still the isomorphism between Pauli matrices and Clifford algebra $A(Si)$ is well established at any order.

Passing from the algebra A(Si) to $N_{i,\pm 1}$ it happens an interesting feature. Consider the case, as example, of $e_3$. While in $A(Si)$ $e_3$ is an abstract algebraic element that has the potentiality to assume or the value +1 or the value –1 (in correspondence, in quantum mechanics it is an operator with possible eigenvalues $\pm 1$), when we pass in the algebra $N_{i,\pm 1}$, $e_3$ is no more an abstract element in this algebra, it becomes a parameter to which we may attribute the numerical value +1, and we have $N_{i,+1}$ whose three abstract element now are $(e_1, e_2, i)$ with commutation rules given in the (2.14). If we attribute to $e_3$ the numerical value -1, we are in $N_{i,-1}$ whose three abstract elements are still $(e_1, e_2, i)$, and the commutation rules are given in (2.17). Reading this statement in the language and in the framework of a quantum mechanical measurement, it means that if we are measuring the given quantum system S with a measuring apparatus and, as result of the actualized and performed measurement, we read the result +1, we are in the corresponding algebraic case, in the algebra $N_{i,+1}$. If instead, performing the measurement, we read the result –1, in this case we are in the algebra $N_{i,-1}$. In each of the two cases this means that a collapse of the wave function has happened.

During a process of quantum measurement, speaking in terms of Clifford algebraic framework, we could have the passage from the Clifford algebra A(Si), in the case in which the result of the measurement of $e_3$ is +1 (read on the instrument), and instead we could have the passage to the new $N_{i,-1}$ Clifford algebra, in the case in which the result of the quantum measurement of $e_3$ gives value –1 (read on the instrument).

In such way it seems that a reformulation of von Neumann's projection postulate may be suggested. The reformulation is that, during a quantum measurement (wave-function collapse), we have the passage from the Clifford algebra A(Si), to the new Clifford algebra $N_{i,\pm 1}$. In brief :

Quantum Measurement (wave-function collapse) $=$ passage from algebra $A(Si)$ to $N_{i,\pm 1}$.

In conclusion we think that the two previously shown theorems in Clifford algebraic framework give justification of the von Neumann's projection postulate and they seem to suggest, in addition, that we may use the passage from the algebra A(Si) to $N_{i,\pm 1}$ to describe actually performed quantum measurements.

A detailed exposition of such results has been discussed by us in ref. [9] but we may discuss still here some illustrative examples.

Let us start discussing a preliminary application.

Assume a two –level microscopic quantum system S with two states $u_+$, $u_-$ corresponding to energy eigenvalues $\varepsilon_+$, $\varepsilon_-$. The Hamiltonian operator $H_S$ can be written

$$H_S = \frac{1}{2}\varepsilon_+(1+e_3) + \frac{1}{2}\varepsilon_-(1-e_3) = \frac{1}{2}(\varepsilon_+ + \varepsilon_-) + \frac{1}{2}(\varepsilon_+ - \varepsilon_-)e_3 \qquad (2.18)$$

In the standard quantum methodological approach we have that

$$u_+ = \begin{pmatrix} 1 \\ 0 \end{pmatrix} \quad , \quad u_- = \begin{pmatrix} 0 \\ 1 \end{pmatrix} \quad , \quad \text{and} \quad H_S u_i = \varepsilon_i u_i . \qquad (2.19)$$

We may also choose $\varepsilon_+ = \varepsilon$ and $\varepsilon_- = 0$ simplifying the (2.18) to

$$H_S = \frac{1}{2}(1+e_3)\varepsilon \qquad (2.20)$$

Indicate an arbitrary state of such quantum microsystem as

$$\psi_S = c_+ u_+ + c_- u_- \qquad (2.21)$$

where, according to Born's rule, we have

$$c_+ = \sqrt{p_+} e^{i\delta_1} \quad , \quad c_- = \sqrt{p_-} e^{i\delta_2} \qquad (2.22)$$

$$p_j \quad (j = +,-) \qquad (2.23)$$

corresponding probabilities with $p_+ + p_- = 1$.

This is the standard quantum mechanical formulation of the system.

Let us admit now that we want to measure the energy of $S$ using a proper apparatus . The rules of quantum mechanics tell us that we will obtain the value $\varepsilon$ with probability $p_+$, and the value zero with probability $p_-$. After the measurement the state of $S$ will be either $u_+$ or $u_-$ according to the measured value of the energy. The experiment will enable us also to estimate $p_+$ as well as $p_-$.

In such simple quantum mechanical example we have , as known, the (2.18)), $e_3$, the (2.20) that are linear Hermitean operators with quantum states acting on the proper Hilbert space.

Let us see instead the question from our Clifford algebraic point of view.

The $e_3$, and $H_S$ given in the (2.18) or in the (2.20) are members of the $A(Si)$ Clifford algebra with basic rules $e_1^2 = e_2^2 = e_3^2 = 1$

$$e_1e_2 = -e_2e_1 = ie_3; \;\; e_2e_3 = -e_3e_2 = ie_1 \;\; ; \;\; e_2e_3 = -e_3e_2 = ie_1; \;\; i = e_1e_2e_3 \qquad (2.24)$$

However, on the basis of theorems n.1 and n.2 shown in the previous sections, starting with the Clifford algebra A(Si), we must use the existing Clifford, dihedral algebra, $N_{i,\pm 1}$ when we arrive to attribute (by a measurement) as example to $e_3$ in one case the numerical value +1 and, in the other case, the numerical value –1.

In the first case we have a dihedral Clifford $N_i$ algebra that is given in the following manner:

$$e_1^2 = e_2^2 = 1 \; i^2 = -1$$

$$e_1e_2 = i, \;\; e_2e_1 = -i, \;\; e_2i = -e_1, \;\; e_2i = -e_1, \;\; ie_2 = e_1, \;\; e_1i = e_2, \;\; ie_1 = -e_2 \qquad (2.25)$$

attributing to $e_3$ the numerical value +1 (in analogy with quantum mechanics: the quantum measurement process has given as result +1). In the second case, we have instead that

$$e_1^2 = e_2^2 = 1; \;\; i^2 = -1;$$

$$e_1e_2 = -i \;, \; e_2e_1 = i, \; e_2i = e_1, ie_2 = -e_1, e_1i = -e_2, \; ie_1 = e_2 \qquad (2.26)$$

that holds when we have arrived to attribute to $e_3$ the numerical value –1 by a direct measurement

Reasoning in terms of a Clifford algebraic framework, we are authorized to apply the passage from algebra A(Si) to algebra $N_{i,\pm 1}$ in the (2.18). From it , we obtain:

$$H_{S(Clifford-element)} = \varepsilon_+ \quad (2.27)$$

if the instrument has given as result of the measurement, the value +1 to $e_3$ (Clifford algebraic parameter of dihedral $N_{i,+1}$ algebra ), and

$$H_{S(Clifford-element)} = \varepsilon_- \quad (2.28)$$

In the first case, we have

$$H_{S(Clifford-element)} = \varepsilon \quad (2.29)$$

and in the second case, we have

$$H_{S(Clifford-element)} = 0 \quad (2.30)$$

Consider now the second application .

Let us introduce a two state quantum system S with connected quantum observable $\sigma_3(e_3)$. We have

$$|\psi\rangle = c_1|\varphi_1\rangle + c_2|\varphi_2\rangle \ , \ \varphi_1 = \begin{pmatrix} 1 \\ 0 \end{pmatrix} \ , \ \varphi_2 = \begin{pmatrix} 0 \\ 1 \end{pmatrix} \quad (2.31)$$

and

$$|c_1|^2 + |c_2|^2 = 1$$

As we know, the density matrix of such system is easily written

$$\rho = a + be_1 + ce_2 + de_3 \quad (2.32)$$

with

$$a = \frac{|c_1|^2 + |c_2|^2}{2} \ , \ b = \frac{c_1^* c_2 + c_1 c_2^*}{2} \ , \ c = \frac{i(c_1 c_2^* - c_1^* c_2)}{2}, \ d = \frac{|c_1|^2 - |c_2|^2}{2} \quad (2.33)$$

where in matrix notation, $e_1$ , $e_2$, and $e_3$ are the well known Pauli matrices

$$e_1 = \begin{pmatrix} 0 & 1 \\ 1 & 0 \end{pmatrix} \ , \ e_2 = \begin{pmatrix} 0 & -i \\ i & 0 \end{pmatrix} \ , \ e_3 = \begin{pmatrix} 1 & 0 \\ 0 & -1 \end{pmatrix} \quad (2.34)$$

Of course, the analogy still holds. The (2.32) is still an element of the $A(Si)$ Clifford algebra. As Clifford algebraic member, the (2.32) satisfies the requirement to be $\rho^2 = \rho$ and $\mathrm{Tr}(\rho) = 1$ under the conditions $a = 1/2$ and $a^2 - b^2 - c^2 - d^2 = 0$ as we evidenced in the (2.4) . In the algebraic framework, let us admit that we attribute to $e_3$ the value +1 (that is to say … the quantum observable $\sigma_3$ assumes the value +1 during quantum measurement ) or to $e_3$ the numerical value –1 (that is to say… the quantum observable $\sigma_3$ assumes the value –1 during the quantum measurement). As previously shown, in such two cases the algebra A, (Si) no more holds, and it will be replaced from the Clifford $N_{i,\pm 1}$. To examine the consequences, starting with the algebraic element (2.32), write it in the two equivalent algebraic forms that are obviously still in the algebra A(Si).

$$\rho = \frac{1}{2}(|c_1|^2 + |c_2|^2) + \frac{1}{2}(c_1 c_2^*)(e_1 + e_2 i) + \frac{1}{2}(c_1^* c_2)(e_1 - ie_2) + \frac{1}{2}(|c_1|^2 - |c_2|^2)e_3 \quad (2.35)$$

and

$$\rho = \frac{1}{2}(|c_1|^2 + |c_2|^2) + \frac{1}{2}(c_1 c_2^*)(e_1 + ie_2) + \frac{1}{2}(c_1^* c_2)(e_1 - e_2 i) + \frac{1}{2}(|c_1|^2 - |c_2|^2)e_3 \quad (2.36)$$

Both such expressions contain the following interference terms.

$$\frac{1}{2}(c_1 c_2^*)(e_1 + e_2 i) + \frac{1}{2}(c_1^* c_2)(e_1 - ie_2) \quad (2.37)$$

and

$$\frac{1}{2}(c_1 c_2^*)(e_1 + ie_2) + \frac{1}{2}(c_1^* c_2)(e_1 - e_2 i) \quad (2.38)$$

Let us consider now that the quantum measurement gives as result +1 for $e_3$. In this case there are the (2.35) and the (2.37) that we must take in consideration. On the basis of our principle, we know that the previous Clifford algebra A(Si) no more holds, but instead it is valid the $N_{1,+1}$ that has the following new commutation rules:

$$e_1 e_2 = i \ , \ e_2 e_1 = -i \ , \ e_2 i = -e_1 \ , ie_2 = e_1 \ , e_1 i = e_2 \ , \ ie_1 = -e_2 \quad (2.39)$$

Inserting such new commutation rules in the (2.35) and in the (2.36), the interference terms are erased and the density matrix, given in the (2.35), now becomes

$$\rho \to \rho_M = |c_1|^2 \quad (2.40)$$

The collapse has happened.

In the same manner let us consider instead that the quantum measurement gives as result -1 for $e_3$. In this case there are the (2.36) and the (2.38) that we take in consideration The Clifford algebra $A(Si)$ no more holds, but instead it is valid the $N_{1,-1}$ that has the following new commutation rules

$$e_1 e_2 = -i \ , \ e_2 e_1 = i \ , \ e_2 i = e_1 \ , ie_2 = -e_1 \ , e_1 i = -e_2 \ , \ ie_1 = e_2 \quad (2.41)$$

Inserting such new commutation rules in the (2.36) and (2.38), remembering that the parameter $e_3$ now assumes value –1, one sees that the interference terms are erased and the density matrix now becomes

$$\rho \to \rho_M = |c_2|^2 \quad (2.42)$$

The collapse has happened.

By using the Clifford bare bone skeleton , we conclude that quantum mechanics now becomes a self-consistent theory since by the $A(Si)$ and $N_{i,\pm 1}$ algebras, the formulation becomes able to describe the collapse of the wave function without recovering an outside ad hoc postulate on quantum measurement as initially formulated by von Neumann.

Let us examine in detail von Neumann results.
Consider the spinor basis given in (2.31).
According to such *projection postulate* the complete phase-damping way for a two state system may be written

$$D(\rho) = |0><0|\rho|0><0| \quad + \quad |1><1|\rho|1><1|$$

where the effect of this mapping is to zero-out the off-diagonal entries of a density matrix:

$$D\begin{pmatrix} \alpha & \beta \\ \gamma & \delta \end{pmatrix} = \begin{pmatrix} \alpha & 0 \\ 0 & \delta \end{pmatrix}$$

If we have a set of mutually orthogonal projection operators ($P_1, P_2, ....., P_m$) which complete to identity, i.e., $P_i P_j = \delta_{ij} P_j$ and $\sum_i P_i = 1$ when a measurement is carried out on a system with state $|\psi>$ then

(1) The result *i* is obtained with probability $p_i = <\psi|P_i|\psi>$

(2) The state collapses to

$$\frac{1}{\sqrt{p_i}} P_i |\psi>$$

The projection operators are the idempotents in the A(si) Clifford algebra.
We have that

$$|0><0| \quad \text{and} \quad |1><1| \tag{2.43}$$

are respectively the idempotents

$$\frac{1+e_3}{2} \quad \text{and} \quad \frac{1-e_3}{2} \tag{2.44}$$

We have that

$$(\frac{1+e_3}{2})\rho(\frac{1+e_3}{2}) \tag{2.45}$$

that gives

$$(\frac{1+e_3}{2})\rho(\frac{1+e_3}{2}) = \alpha\ (\frac{1+e_3}{2}) \tag{2.46}$$

and explicitly

$$\begin{pmatrix} \alpha & 0 \\ 0 & 0 \end{pmatrix} \tag{2.47}$$

In the case of

$$\frac{1-e_3}{2} \tag{2.48}$$

one obtains as result

$$\beta(\frac{1-e_3}{2}) \tag{2.49}$$

and explicitly

$$\begin{pmatrix} 0 & 0 \\ 0 & \delta \end{pmatrix}$$

The sum gives

$$\begin{pmatrix} \alpha & 0 \\ 0 & \delta \end{pmatrix}$$

Generally speaking, given an observable with connected linear Hermitean operator $O$ having eigenvalues $O_1, O_2, \ldots\ldots\ldots$
we have

$$\text{Prob}.(O_n) = Tr(P_n \rho) \quad (2.49a)$$

that obviously is fully justified by our $N_{i,\pm 1}$ theorem.

In conclusion we have given a full Clifford algebraic justification of von Neumann's projection postulate.

Note that we have involved idempotents in the $A(Si)$ Clifford algebraic quantum scheme, and they have projectors as counterpart in standard quantum physics. We cannot ignore a fundamental step: according to J. von Neumann projection operators represent logical statements [6]. We have verified that they assume the same meaning in our algebraic scheme. Consequently we cannot escape to the conclusion previously introduced. Measurements must be intended as semantic acts, and conceptual entities are represented in our bare bone skeleton of quantum mechanics as a motor device as well as objects and matter dynamics.

Obviously we may describe the wave function collapse also using a time dependent formalism. The elaboration has been exposed in detail by us elsewhere [9] . We will reassume it here.

Consider the quantum system S and indicate by $\psi_0$ the state at the initial time 0. The state at any time t will be given by

$$\psi(t) = U(t)\psi_0 \quad \text{and} \quad \psi_0 = \psi(t=0) \quad (2.50)$$

An Hamiltonian H must be constructed such that the evolution operator U(t), that must be unitary, gives $U(t) = e^{-iHt}$ .

It is well known that, given a finite N-level quantum system described by the state $\psi$, its evolution is regulated according to the time dependent Schrödinger equation

$$i\hbar \frac{d\psi(t)}{dt} = H(t)\psi(t) \quad \text{with } \psi(0) = \psi_0 . \quad (2.51)$$

Let us introduce a model for the hamiltonian H(t). We indicate by $H_0$ the hamiltonian of the system S, and we add to $H_0$ an external time varying hamiltonian, $H_1(t)$, representing the perturbation to which the system S is subjected by action of the measuring apparatus. We write the total hamiltonian as

$$H(t) = H_0 + H_1(t) \quad (2.52)$$

so that the time evolution will be given by the following Schrödinger equation

$$i\hbar\frac{d\psi(t)}{dt}=[H_0+H_1(t)]\psi(t) \qquad (2.52a)$$

We have that

$$i\hbar\frac{dU(t)}{dt}=H(t)U(t)=[H_0+H_1(t)]U(t) \quad \text{and } U(0)=I \qquad (2.53)$$

where U(t) pertains to the special group SU(N).

Let $A_1, A_2, \ldots\ldots, A_n$ , ($n=N^2-1$), are skew-hermitean matrices forming a basis of Lie algebra SU(N). In this manner one arrives to write the explicit expression of the hamiltonian H(t). It is given in the following manner

$$-iH(t)=-i[H_0+H_1(t)]=\sum_{j=1}^{n}a_jA_j+\sum_{j=1}^{n}b_jA_j \qquad (2.54)$$

where $a_j$ and $b_j = b_j(t)$ are respectively the constant components of the free hamiltonian and the time-varying control parameters characterizing the action of the measuring apparatus , just the semantic act.. If we introduce T, the time ordering parameter (for details see reff. [9,10,11]), we have

$$U(t)=T\exp(-i\int_0^t H(\tau)d\tau)=T\exp(-i\int_0^t (a_j+b_j(\tau))A_jd\tau) \qquad (2.55)$$

that is the well known Magnus expansion. Locally U(t) may be expressed by exponential terms as it follows

$$U(t)=\exp(\gamma_1A_1+\gamma_2A_2+\ldots\ldots+\gamma_nA_n) \qquad (2.56)$$

on the basis of the Wein-Norman formula

$$\Xi(\gamma_1,\gamma_2,\ldots\ldots,\gamma_n)\begin{pmatrix}\dot\gamma_1\\ \dot\gamma_2\\ \ldots\\ \dot\gamma_n\end{pmatrix}=\begin{pmatrix}a_1+b_1\\ a_2+b_2\\ \ldots\\ a_n+b_n\end{pmatrix} \qquad (2.57)$$

with $\Xi$ n x n matrix, analytic in the variables $\gamma_i$. We have $\gamma_i(0)=0$ and $\Xi(0)=I$, and thus it is invertible. We obtain

$$\begin{pmatrix}\dot\gamma_1\\ \dot\gamma_2\\ \ldots\\ \dot\gamma_n\end{pmatrix}=\Xi^{-1}\begin{pmatrix}a_1+b_1\\ a_2+b_2\\ \ldots\\ a_n+b_n\end{pmatrix} \qquad (2.58)$$

Consider a simple case based on the superposition of only two states. We have

$$\psi=[y_1,y_2]^T \quad \text{and} \quad |y_1|^2+|y_2|^2=1 \qquad (2.59)$$

We have here an SU(2) unitary transformation, selecting the skew symmetric basis for SU(2). We will have that

$$e_1 = \begin{pmatrix} 0 & 1 \\ 1 & 0 \end{pmatrix} \ , \ e_2 = \begin{pmatrix} 0 & -i \\ i & 0 \end{pmatrix} \ , \ e_3 = \begin{pmatrix} 1 & 0 \\ 0 & -1 \end{pmatrix} \tag{2.60}$$

The following matrices are given

$$A_j = \frac{i}{2} e_j \ , \mathrm{j} = 1,2,3 \tag{2.61}$$

The reader may now ascertain that the previously developed formalism is moving in direct correspondence with our Clifford algebra A(Si).

We are now in the condition to express H(t) and U(t) in our case of interest. The most simple situation we may examine is that one of fixed and constant control parameters $b_j$. The hamiltonian H will become fully linear time invariant and its exponential solution will take the following form

$$e^{t(\sum_{j=1}^{3}(a_j+b_j)A_j)} = \cos(\frac{k}{2}t)I + \frac{2}{k}sen(\frac{k}{2}t)\left(\sum_{j=1}^{3}(a_j + b_j)A_j\right) \tag{2.62}$$

with $k = \sqrt{(a_1+b_1)^2 + (a_2+b_2)^2 + (a_3+b_3)^2}$ . In matrix form it will result

$$U(t) = \begin{pmatrix} \cos\frac{k}{2}t + \frac{i}{k} sen\frac{k}{2}t(a_3+b_3) & \frac{1}{k} sen\frac{k}{2}t[a_2+b_2+i(a_1+b_1)] \\ \frac{1}{k} sen\frac{k}{2}t[-a_2-b_2+i(a_1+b_1)] & \cos\frac{k}{2}t - \frac{i}{k} sen\frac{k}{2}t(a_3+b_3) \end{pmatrix} \tag{2.63}$$

and, obviously, it will result to be unimodular as required.

Starting with this matrix representation of time evolution operator U(t), we may deduce promptly the dynamic time evolution of quantum state at any time t writing

$$\psi(t) = U(t)\psi_0 \tag{2.64}$$

assuming that we have for $\psi_0$ the following expression

$$\psi_0 = \begin{pmatrix} c_{true} \\ c_{false} \end{pmatrix} \tag{2.65}$$

having adopted for the true and false states (or yes-not states, +1 and –1 corresponding eigenvalues of such states) the following matrix expressions

$\varphi_{true} = \begin{pmatrix} 1 \\ 0 \end{pmatrix}$ and $\varphi_{false} = \begin{pmatrix} 0 \\ 1 \end{pmatrix}$

Finally, one obtains the expression of the state $\psi(t)$ at any time

$$\psi(t) = \left[c_{true}\left[\cos\frac{k}{2}t + \frac{i}{k}sen\frac{k}{2}t(a_3+b_3)\right] + c_{false}\left[\frac{1}{k}sen\frac{k}{2}t[(a_2+b_2)+i(a_1+b_1)]\right]\right]\varphi_{true} +$$

$$\left[c_{true}\left[\frac{1}{k}sen\frac{k}{2}t[i(a_1+b_1)-(a_2+b_2)]\right] + c_{false}\left[\cos\frac{k}{2}t - \frac{i}{k}sen\frac{k}{2}t(a_3+b_3)\right]\right]\varphi_{false} \tag{2.66}$$

As consequence, the two probabilities $P_{true}(t)$ and $P_{false}(t)$, will be given at any time t by the following expressions

$$P_{true}(t) = (A^2 + B^2)\cos^2\frac{k}{2}t + \frac{1}{k^2}sen^2\frac{k}{2}t(P^2 + Q^2) + \frac{senkt}{k}(AP + BQ) \qquad (2.67)$$

and

$$P_{false}(t) = (C^2 + D^2)\cos^2\frac{k}{2}t + \frac{1}{k^2}sen^2\frac{k}{2}t(S^2 + R^2) + \frac{senkt}{k}(RC + DS) \qquad (2.68)$$

where

A= Re $c_{true}$ , B=Im $c_{true}$, C=Re $c_{false}$ , D=Im $c_{false}$ ,

$P=-D(a_1+b_1)+C(a_2+b_2)-B(a_3+b_3)$,

$Q=C(a_1+b_1)+D(a_2+b_2)+A(a_3+b_3)$,

$R=-B(a_1+b_1)-A(a_2+b_2)+D(a_3+b_3)$,

$S=A(a_1+b_1)-B(a_2+b_2)-C(a_3+b_3)$

Until here we have developed only standard quantum mechanics. The reason to have developed here such formalism has been to evidence that at each step it has its corresponding counterpart in Clifford algebraic framework A(Si), and thus we may apply to it the two theorems previously demonstrated, passing from the algebra $A(Si)$ to $N_{i,\pm 1}$. In fact, to this purpose, it is sufficient to multiply the (2.63) by the (2.65) to obtain the final forms of $c_{true}(t)$ and $c_{false}(t)$

In the final state we have that

$$\psi_t = \begin{pmatrix} c_{true}(t) \\ c_{false}(t) \end{pmatrix} \qquad (2.69)$$

We may now write the density matrix that will result to have the same structure of the previously case given in the (2.32) but obviously with explicit evidence of time dependence. In the Clifford algebraic framework it will pertain still to the Clifford algebra A( Si). In order to describe the wave-function collapse we have to repeat the same procedure that we developed previously from the (2.32) to the (2.42), considering that, in accord to our criterium, we have to pass from the algebra A(Si) to $N_{i,\pm 1}$, and obtaining

$$\rho \to \rho_M = \left|c_{true}(t)\right|^2 \qquad (2.70)$$

in the case $N_{i,+1}$

and

$$\rho \to \rho_M = \left|c_{false}(t)\right|^2 \qquad (2.71)$$

in the case $N_{i,-1}$, as required in the collapse.

Note that, using Clifford algebra, we have given now a complete theoretical elaboration of the problem of wave function reduction in quantum mechanics also considering the process under the profile of the time dynamics.

Evidences of such elaboration have been given by us at cognitive level using introducing also experimental verifications [11].

There is still another feature that is necessary to explain, and we will develop it here now.

Until here we considered only examples of two state quantum systems. Let us expand our formulation introducing the Clifford algebra at any order n.
First consider Clifford $A(Si)$ algebra at order n=4 (for details see our previous papers and references therein). One has
$E_{0\,i} = I^{1} \otimes e_{i};\qquad E_{i\,0} = e_{i} \otimes I^{2}$ (2.72a)

The notation ⊗ denotes direct product of matrices, and $I^{i}$ is the $i$th 2x2 unit matrix. Thus, in the case of $n= 4$ we have two distinct sets of Clifford basic unities, $E_{0\,i}$ and $E_{i\,0}$, with
$E^{2}_{0i} = 1;\qquad E^{2}_{i0} = 1,\qquad i = 1, 2, 3;$ (2.72b)

$E_{0\,i}E_{0\,j} = i\,E_{0\,k};\qquad E_{i\,0}E_{j\,0} = i\,E_{k\,0},\qquad j = 1, 2, 3;\qquad i \neq j$
and
$E_{i0}\,E_{0\,j} = E_{0\,j}\,E_{i\,0}$ (2.73)

with $(i, j, k)$ cyclic permutation of $(1, 2, 3)$.
Let us examine now the following result
$(I^{1} \otimes e_{i})\,(e_{j} \otimes I^{2}) = E_{0\,i}\,E_{j\,0} = E_{j\,i}$ (2.74)

It is obtained according to our basic rule on cyclic permutation required for Clifford basic unities. We have that $E_{0\,i}\,E_{j0} = E_{j\,i}$ with $i = 1, 2, 3$ and $j=1, 2, 3,$ with $E^{2}_{j\,i} = 1$, $E_{i\,j}\,E_{k\,m} \neq E_{k\,m}\,E_{i\,j}$, and $E_{i\,j}\,E_{k\,m} = E_{p\,q}$ where $p$ results from the cyclic permutation $(i, k, p)$ of $(1, 2, 3)$ and $q$ results from the cyclic permutation $(j, m, q)$ of $(1, 2, 3)$.
In the case $n = 4$ we have two distinct basic set of unities $E_{0\,i}$, $E_{i\,0}$ and, in addition, basic sets of unities $(E_{i\,j}, E_{i\,p}, E_{0\,m})$ with $(j, p, m)$ basic permutation of $(1, 2, 3)$.
This is the Clifford algebra A at order n=4.
In the other more general cases we have $E_{0\,0\,i}$, $E_{0\,i\,0}$, and $E_{i\,0\,0}$, $i = 1, 2, 3$ and

$E_{0\,0\,i} = I^{1} \otimes I^{1} \otimes e_{i};\quad E_{0\,i0} = I^{2} \otimes e_{i} \otimes I^{2};\qquad E_{i\,0\,0} = e_{i} \otimes I^{3} \otimes I^{3}$
and
$(I^{1} \otimes I^{1} \otimes e_{i}) \cdot (I^{2} \otimes e_{i} \otimes I^{2}) \cdot (e_{i} \otimes I^{3} \otimes I^{3}) = e_{i} \otimes e_{i} \otimes e_{i} =$
$= E_{0\,0\,i}\,E_{0\,i\,0}\,E_{i\,0\,0} = E_{i\,i\,i}$ (2.75)
Still we will have that
$E_{0\,0\,i}\,E_{0\,i\,0} = E_{0\,i\,0}\,E_{i\,0\,0};\quad E_{0\,0\,i}E_{i\,0\,0} = E_{i\,0\,0}E_{0\,0\,i};\qquad E_{0\,i\,0}\,E_{i\,0\,0} = E_{i\,0\,0}E_{0\,i0}$

Generally speaking, fixed the order $n$ of the $A(Si)$ Clifford algebra in consideration , we will have that

$$\Gamma_1 = \Lambda_n$$

$$\Gamma_{2m} = \Lambda_{n-m} \otimes e_2^{(n-m+1)} \otimes I^{(n-m+2)} \otimes \ldots\ldots\ldots \otimes I^n$$

$$\Gamma_{2m+1} = \Lambda_{n-m} \otimes e_3^{(n-m+1)} \otimes I^{(n-m+2)} \otimes \ldots\ldots\ldots \otimes I^n \qquad (2.75b)$$

$$\Gamma_{2n} = e_2 \otimes I^{(2)} \otimes \ldots\ldots\ldots \otimes I^n$$

with

$$\Lambda_n = e_1^{(1)} \otimes e_1^{(2)} \otimes \ldots\ldots \otimes e_1^{(n)} = (e_1 \otimes I^{(1)} \otimes \ldots\ldots \otimes I^n)\cdot(\ldots\ldots\ldots)\cdot(I^{(1)} \otimes I^{(2)} \ldots \otimes I^{(n)} \otimes e_1);$$

$m = 1, \ldots\ldots, n-1$

according to the $n$-possible dispositions of $e_1$ and $I^1, I^2, \ldots, I^n$ in the distinct direct products.

We may now give the explicit expressions of $E_{0i}$, $E_{i0}$, and $E_{ij}$ at the order n=4.

$$E_{01} = \begin{pmatrix} 0 & 1 & 0 & 0 \\ 1 & 0 & 0 & 0 \\ 0 & 0 & 0 & 1 \\ 0 & 0 & 1 & 0 \end{pmatrix}; \; E_{02} = \begin{pmatrix} 0 & -i & 0 & 0 \\ i & 0 & 0 & 0 \\ 0 & 0 & 0 & -i \\ 0 & 0 & i & 0 \end{pmatrix}; \; E_{03} = \begin{pmatrix} 1 & 0 & 0 & 0 \\ 0 & -1 & 0 & 0 \\ 0 & 0 & 1 & 0 \\ 0 & 0 & 0 & -1 \end{pmatrix} \qquad (2.76)$$

$$E_{10} = \begin{pmatrix} 0 & 0 & 1 & 0 \\ 0 & 0 & 0 & 1 \\ 1 & 0 & 0 & 0 \\ 0 & 1 & 0 & 0 \end{pmatrix}; \; E_{20} = \begin{pmatrix} 0 & 0 & -i & 0 \\ 0 & 0 & 0 & -i \\ i & 0 & 0 & 0 \\ 0 & i & 0 & 0 \end{pmatrix}; E_{30} = \begin{pmatrix} 1 & 0 & 0 & 0 \\ 0 & 1 & 0 & 0 \\ 0 & 0 & -1 & 0 \\ 0 & 0 & 0 & -1 \end{pmatrix};$$

$$E_{11} = \begin{pmatrix} 0 & 0 & 0 & 1 \\ 0 & 0 & 1 & 0 \\ 0 & 1 & 0 & 0 \\ 1 & 0 & 0 & 0 \end{pmatrix}; \; E_{22} = \begin{pmatrix} 0 & 0 & 0 & -1 \\ 0 & 0 & 1 & 0 \\ 0 & 1 & 0 & 0 \\ -1 & 0 & 0 & 0 \end{pmatrix}; \; E_{33} = \begin{pmatrix} 1 & 0 & 0 & 0 \\ 0 & -1 & 0 & 0 \\ 0 & 0 & -1 & 0 \\ 0 & 0 & 0 & 1 \end{pmatrix};$$

$$E_{12} = \begin{pmatrix} 0 & 0 & 0 & -i \\ 0 & 0 & i & 0 \\ 0 & -i & 0 & 0 \\ i & 0 & 0 & 0 \end{pmatrix}; E_{13} = \begin{pmatrix} 0 & 0 & 1 & 0 \\ 0 & 0 & 0 & -1 \\ 1 & 0 & 0 & 0 \\ 0 & -1 & 0 & 0 \end{pmatrix}; E_{21} = \begin{pmatrix} 0 & 0 & 0 & -i \\ 0 & 0 & -i & 0 \\ 0 & i & 0 & 0 \\ i & 0 & 0 & 0 \end{pmatrix};$$

$$E_{31}=\begin{pmatrix}0&1&0&0\\1&0&0&0\\0&0&0&-1\\0&0&-1&0\end{pmatrix};E_{23}=\begin{pmatrix}0&0&-i&0\\0&0&0&i\\i&0&0&0\\0&-i&0&0\end{pmatrix};E_{32}=\begin{pmatrix}0&-i&0&0\\i&0&0&0\\0&0&0&i\\0&0&-i&0\end{pmatrix}.$$

Note the following basic feature: we have now some different sets of Clifford algebras $A(Si)$. In detail, we have the following sets of basic $A(Si)$ Clifford algebras:

$$\begin{aligned}&(E_{01},E_{12},E_{13}),(E_{01},E_{22},E_{23}),(E_{01},E_{32},E_{33}),(E_{02},E_{11},E_{13}),\\&(E_{02},E_{21},E_{23}),(E_{02},E_{31},E_{33}),(E_{03},E_{11},E_{12}),\\&(E_{03},E_{21},E_{22}),(E_{03},E_{31},E_{32}),(E_{10},E_{23},E_{33}),(E_{10},E_{22},E_{32}),\\&(E_{10},E_{21},E_{31}),(E_{20},E_{13},E_{33}),(E_{20},E_{12},E_{32}),(E_{20},E_{11},E_{31}),\\&(E_{30},E_{13},E_{23}),(E_{30},E_{12},E_{22}),(E_{30},E_{11},E_{21})\end{aligned} \quad (2.77)$$

To each of these sets we may apply the theorems n.1 and n.2 previously shown and we may also apply the criterium of the passage from the $A(Si)$ to the $N_{1,\pm1}$ that we have just used in the other previous cases of application.

Fixed such algebraic features, consider the problem that is often formulated in standard quantum mechanics. It is that, in order to avoid possible contradictions, we should still modify the previous elaboration for the wave-function collapse, by introducing the states of a given measurement apparatus system A obtaining in this case

$$\rho=\rho_S\otimes\rho_A=\sum_i\sum_j c_i c_j^*|\varphi_i><\varphi_j|\otimes\rho_A\to\rho_{S,A,t}=\sum_k|c_k|^2|\varphi_k><\varphi_k|_t\otimes\rho_{A(k),t} \quad (2.78)$$

We may refer the algebraic sets $E_{0i}$ to the quantum system S to be measured, and consider the algebraic sets $E_{i0}$ to the measuring apparatus A. Still we have the basic algebraic set $E_{ij}$ that relates the coupling of S with A. Let us write the density matrix $\rho$ at such order n=4. To simplify, we may write it in the following general form

$$\rho=\begin{pmatrix}a&b_1+ib_2&c_1+ic_2&d_1+id_2\\b_1-ib_2&e&f_1+if_2&q_1+iq_2\\c_1-ic_2&f_1-if_2&h&t_1+it_2\\d_1-id_2&q_1-iq_2&t_1-it_2&s\end{pmatrix} \quad (2.79)$$

Obviously, the correspondence between Clifford algebra and quantum mechanics still holds also at the present order. The $\rho$ of the (2.79) is still a member of the Clifford algebra A(Si) that in fact, on the basis of the (2.76) may be written in the following manner

$$\rho = a\left(\frac{E_{00}+E_{03}+E_{30}+E_{33}}{4}\right)+e\left(\frac{E_{00}+E_{30}-E_{03}-E_{33}}{4}\right)+$$
$$h\left(\frac{E_{00}+E_{03}-E_{30}-E_{33}}{4}\right)+s\left(\frac{E_{00}-E_{03}-E_{30}+E_{33}}{4}\right)+$$
$$\left[b_1\left(\frac{E_{01}+E_{31}}{2}\right)-b_2\left(\frac{E_{02}+E_{32}}{2}\right)\right]+\left[c_1\left(\frac{E_{10}+E_{13}}{2}\right)-c_2\left(\frac{E_{23}+E_{20}}{2}\right)\right]+$$
$$\left[d_1\left(\frac{E_{11}-E_{22}}{2}\right)-d_2\left(\frac{E_{12}+E_{21}}{2}\right)\right]+\left[f_1\left(\frac{E_{11}+E_{22}}{2}\right)+f_2\left(\frac{E_{12}-E_{21}}{2}\right)\right]+$$
$$\left[q_1\left(\frac{E_{10}-E_{13}}{2}\right)+q_2\left(\frac{E_{23}-E_{20}}{2}\right)\right]+\left[t_1\left(\frac{E_{01}-E_{31}}{2}\right)+t_2\left(\frac{E_{32}-E_{02}}{2}\right)\right] \quad (2.80)$$

It is in $A(Si)$ Applying the previous criterium relating the quantum measurement, we must now pass from A(Si) to $N_{i,\pm 1}$. Let us start considering for $E_{33}$ the numerical value +1 and this is to say that or $E_{03}=E_{30}=+1$ or $E_{03}=E_{30}=-1$.
On the basis of such condition of the measuring instrument, by inspection of the (2.80) , it is seen that the terms by $e$ and $h$ go to zero. It remains the term by $a$ for $E_{03}=E_{30}=+1$ and the term in $s$ for $E_{03}=E_{30}=-1$. All the other terms containing $b_i, c_i, d_i, f_i, q_i, t_i$ ($i=1,2$) go to zero and the wave function collapse has happened.
Let us explain as example as the term

$$\frac{E_{02}+E_{32}}{2} \quad (2.81)$$

pertaining to $b_2$, goes to zero.
Remember that we have attributed to $E_{33}$ the value +1. By inspection of the (2.77), it is seen that the basic algebraic set $A(Si)$ in which $E_{33}$ enters is ($E_{01}, E_{32}, E_{33}$).
Passing from the algebra $A(Si)$ to the algebra $N_{i,+1}$ (in fact we have attributed to $E_{33}$ the numerical value +1) we obtain the new commutation rule that

$$E_{01}E_{32}=i\,. \quad (2.82)$$

On the other hand, considering the basic algebraic A(Si) set ($E_{01}, E_{02}, E_{03}$) of the (2.77) with attribution to $E_{03}$ the numerical value -1, we have the new commutation rule that

$$E_{01}E_{02}=-i \quad (2.83)$$

In conclusion we have that

$$E_{32}=E_{01}i \quad (2.84)$$

and

$$\frac{E_{02}+E_{32}}{2}=\frac{E_{02}+E_{01}i}{2}=\frac{-E_{01}i+E_{01}i}{2}=0 \qquad (2.85)$$

Following the same procedure , one obtains that also the other interference terms are erased and in conclusion, passing from the algebra A (Si) to the $N_{i,\pm 1}$, one obtains a substantial equivalence with von Neumann projection postulate. On the other hand the density matrix $\rho$ , given in (2.80) , has been reduced to be

$$\rho = a(\frac{E_{00}+E_{03}+E_{30}+E_{33}}{4})+s(\frac{E_{00}-E_{03}-E_{30}+E_{33}}{4}) \qquad (2.86)$$

where in the new application of the $N_{i,\pm 1}$ algebra , we may have

or

$$E_{03}=E_{30}=+1\ (E_{33}=+1) \qquad (2.87)$$

and thus

$$\rho \to \rho_M = a \qquad (2.88)$$

or

$$E_{03}=E_{30}=-1\ (E_{33}=+1) \qquad (2.89)$$

and thus

$$\rho \to \rho_M = s \qquad (2.90)$$

and the collapse has happened

We have now completed our exposition on the algebraic Clifford formulation of wave function collapse in quantum mechanics.

## 3. On Some Advanced Arguments of Quantum Mechanics.

## 3.1 Derivation of Kochen- Specker theorem by using Clifford algebra

According to various authors , and in particular to A. Peres [12], there is a quantum analog in quantum mechanics as it is usually called the necessity of contextual measurements in quantum mechanics. Of course this is an argument that we have developed in detail elsewhere [10] also giving experimental evidences at cognitive level in humans [11]

In usual quantum mechanical terms one arrives to conclude that we cannot admit context independent actualisations.

A. Peres states [12] :

*The Kochen-Specker theorem, given in 1967 [13], is of fundamental importance for quantum theory. It asserts that, in a Hilbert space of dimension* $d \geq 3$*, it is impossible to associate definite values, 1 or 0, with every projection operator* $P_m$ *in such a way that, if a set of commuting* $P_m$ *satisfies* $\sum P_m = 1$*, the corresponding values* $v(P_m)$ *will also satisfy* $\sum v(P_m) = 1$.

In a simple proof of the theorem he used just the operators corresponding to the algebraic elements that we introduced in (2.72 a,b), and in (2.76), and he writes:

*Consider a pair of spin ½ particles in any state. In the square array we have*

$$\begin{matrix} E_{03} & E_{30} & E_{33} \\ E_{10} & E_{01} & E_{11} \\ E_{13} & E_{31} & E_{22} \end{matrix} \qquad (3.1)$$

*each one of the nine operators has eigenvalues* $\pm 1$. *In each row and in each column, the three operators commute, and each operator is the product of the two others, except in the third column, where an extra minus sign is needed.*

$$E_{13}E_{31} = E_{22} \quad and \quad E_{33}E_{11} = -E_{22} \qquad (3.2)$$

*Because of the opposite signs in the (3.2), it is clearly impossible to attribute to the nine elements of the (3.1) numerical values,* $+1$ *or* $-1$, *which would be the results of the measurements of these operators (if these measurements were performed), and which would obey the same multiplication rule as the operators themselves. We have therefore reached a contradiction. This simple proof shows that what we call "the result of a measurement of A, cannot in general depend only on the choice of A and on the system being measured".*

Let us now apply our two theorems n1 and n.2 relating the $A(Si)$ and the $N_{i,\pm 1}$, considering the elements given in the (3.1) as the algebraic Clifford elements that we have previously introduced. Let us consider the different sets of Clifford algebras $A(Si)$ that are determined at order n= 4, and that we have introduced in the (2.77). Looking now at the (3.1), apply the theorem n.2 assuming to attribute to $E_{33}$ one of the numerical possible values $\pm 1$. Identify in the (2.77) the algebraic Clifford sets that contain $E_{33}$ . As consequence of the theorem n.2, the other basic elements of the algebraic set will remain indeterminate in an irreducible manner. They are $E_{01}$, $E_{32}$, $E_{02}$, $E_{31}$, $E_{10}$, $E_{23}$, $E_{20}$, and $E_{13}$. Consequently the theorem is demonstrated since some of such basic elements enter directly in the (3.1) and thus give proof that we cannot admit context independent actualisations.

We may complete our exposition giving another proof of the theorem. In detail, J. Bricmont [14] gave proof of a no hidden variable theorem.

He states

*Let* **A** *be the set of self-adjoint operators on some Hilbert space (which may be taken of dimension four below).*

*Therem 1. There does not exist a map*

$v\colon \boldsymbol{A} \to R$

*such that*

*1)*

$\forall A \in \boldsymbol{A}, v(A) \in (set\ of\ eigenvalues\ of\ A)$

*2)* (3.2a)

$\forall A, B \in \boldsymbol{A},\ with\ [A,B] = 0, v(AB) = v(A)v(B)$

Let us consider our quantum like algebraic framework. We have that

$$E_{01}E_{20}E_{02}E_{10}E_{01}E_{10}E_{02}E_{20} = -1 \quad (3.3)$$

Let us consider still the following basic elements

$$A = E_{01}E_{20},\ B = E_{02}E_{10}, C = E_{01}E_{10}, D = E_{02}E_{20}, X = AB, Y = CD \quad (3.4)$$

Note that

$[A, B] = 0$ that is to say $E_{01}E_{20}E_{02}E_{10} = E_{02}E_{10}E_{01}E_{20}$ (3.5)

$[C, D] = 0$ that is to say $E_{01}E_{10}E_{02}E_{20} = E_{02}E_{20}E_{01}E_{10}$ (3.6)

$[X, Y] = 0$ that is to say $E_{01}E_{20}E_{02}E_{10}E_{01}E_{10}E_{02}E_{20} = E_{01}E_{10}E_{02}E_{20}E_{01}E_{20}E_{02}E_{10}$ (3.7)

The (3.3) may be rewritten as

$$XY = -1 \quad (3.8)$$

We have now that

$$v(XY) = -1 = v(AB)v(CD) = v(A)v(B)v(C)v(D) =$$
$$v(E_{01})v(E_{20})v(E_{02})v(E_{10})v(E_{01})v(E_{10})v(E_{02})v(E_{20}) =$$
$$= v^2(E_{01})v^2(E_{20})v^2(E_{02})v^2(E_{10}) = +1 \quad (3.9)$$

that is a contradiction.

Following this procedure Bricmont [14] , under the profile of quantum mechanics, concludes :

*The non existence of the map v means that measurements are , as one calls them, contextual , i.e. do not reveal preexisting properties of the system , but, in some sense , produce them.*

Under our algebraic profile let us observe $X = AB$ and $Y = CD$ in (3.4). (3.10)

We have

$X = E_{01}E_{20}E_{02}E_{10}$ that is to say $E_{21}E_{12}$. They pertain this time to the set $(E_{21}, E_{12}, E_{33})$ with $E_{21}E_{12} = E_{12}E_{21} = E_{33}$ , $E_{12}E_{33} = E_{33}E_{12} = E_{21}$, $E_{33}E_{21} = E_{21}E_{33} = E_{12}$ (3.11)

This is to say that in (3.11), again as it happened in the previous given proof , we are considering an algebraic set that violate the basic requirements of the prefixed algebraic structure since in this case the basic elements result to be commutative instead of non commutative with two cyclic permutations $(i\ j\ k)$ of $(1\ 2\ 3)$ that are involved rather than one. The basic algebraic products as given for the case n=2 are violated. In absence of the respect of such basic rules , we are out from a quantum like algebraic structure and consequently contradictions are induced.

The same thing happens for $Y$ given in the (3.4).

We have that

$Y = E_{01}E_{10}E_{02}E_{20}$ that is to say $E_{11}E_{22}$. Therefore, it pertains to the set $(E_{11}, E_{22}, E_{33})$

that again violates all the previously mentioned basic quantum like algebraic rules previously discussed. Again we have violated the basic criterion that we have introduced: according to it, we are out from a quantum like algebraic structure every time in which we violate the basic rules of such algebra. Consequently, contradictions are induced.
We may express the case explicitly. The two assumed sets are
$(E_{21}, E_{12}, E_{33})$ and $(E_{11}, E_{22}, E_{33})$.
The following scheme arises

$$\begin{matrix} E_{01} & E_{10} & E_{11} \\ E_{20} & E_{02} & E_{22} \\ E_{21} & E_{12} & E_{33} \end{matrix} \qquad (3.12)$$

where
$E_{10}E_{01}E_{20}E_{02} = iE_{30}iE_{03} = -E_{33}$
and (3.13)
$E_{20}E_{01}E_{10}E_{02} = -iE_{30}iE_{03} = E_{33}$

## 3.2 The Einstein-Podolski-Rosen Paradox Explained by Using the Clifford Algebra.

As it is known, there are several versions of the paradox, starting with the initial EPR formulation of the authors in 1935 [15].
In this paper we will follow the excellent formulation that was given by Asher Peres in 1992 [16].
We quote directly from this article:
*A fundamental issue was raised by Einstein, Podolsky, and Rosen in a classical article entitled "Can quantum mechanical description of physical reality be considered complete?". In that article, the authors define "elements of physical reality" by the following criterion:*
*If, without in any way disturbing the system, we can predict with certainty … the value of a physical quantity, then there exists an element of physical reality corresponding to this physical quantity.*
*The criterion is then applied by EPR to a composite quantum system consisting of two distant particles, with an entangled wave function such that a measurement performed on one of the particles allows to predict with certainty the results of a similar measurement that can (but need not) be performed on the other, distant particle. It then follows from an analysis of these conceptual measurements that more information potentially exists than can be supplied by the wave function, and thus EPR " are forced to conclude that the quantum mechanical description of physical reality given by wave functions is not complete".*
Asher Peres continues:
*The simplest example of such a situation is that of two spin ½ particles, far apart from each other, in a singlet state. With the standard notations of Pauli matrices, we have*
$(E_{01} + E_{10})\psi = 0$

$(E_{02} + E_{20})\psi = 0$ (3.14)

Actually Peres uses the notations $\sigma_{1j}$ *and* $\sigma_{2j}$ $(j = x, y)$ to represent spin operators for particles 1 and 2. We used instead the algebraic notations $E_{0j}$ *and* $E_{j0}$ $(j = 1,2,3)$ that are respectively the algebraic basic elements that we expressed previously.. In this manner $E_{01}, E_{02}, E_{03}$ are the three basic elements relating the spin of the first particle, and $E_{10}, E_{20}, E_{30}$ are those relating the other particle.

Asher Peres continues:

*The first equation in (3.14) asserts that measurements of* $E_{01}$ *and* $E_{10}$*, if performed, shall yield opposite values,* $m_{1x}$ *and* $m_{2x}$*, respectively. Each one of these operators thus corresponds to an " element of reality" because its value can be ascertained , without perturbing in any way the particle to which this operator pertains, by means of a measurement performed on the other particle. The same interpretation can be given to the second equation in (3.14).*

Asher Peres continues:

*…(in) our example of a pair of spin ½ particles in a singlet state, w may define the numerical value of the product* $E_{01}E_{20}$ *as the product of the individual numerical values* $m_{1x}m_{2y}$*. Likewise, the numerical value of* $E_{10}E_{02}$ *is the product* $m_{2x}m_{1y}$*. From the foregoing discussion, these products must be equal; but, on the other hand, they must be opposite, because the singlet state also satisfies*

$(E_{01}E_{20} + E_{10}E_{02})\psi = 0$ (3.15)

*What we have here is no longer a paradox, but an algebraic contradiction.*

We conclude: it is an algebraic contradiction that is at the origin of the paradox.

Our attempt is now to give a quantum like explanation and solution of such algebraic contradiction and paradox.

## 3.2.a The Interpretation and the Solution of the EPR Paradox.

We must now return to use our Clifford algebraic scheme.

One advantage , using such algebraic scheme, is that we may adopt new algebraic elements to represent combined algebraic elements. They are missing in quantum mechanics operator representation.

In fact, we may now introduce such new basic elements that result from their previous combination

$E_{11} = E_{01}E_{10} = E_{10}E_{01}$ ; $E_{22} = E_{02}E_{20} = E_{20}E_{02}$ (3.16)

and

$E_{33} = E_{03}E_{30} = E_{30}E_{03}$ (3.17)

According to the paradox, we attribute to $E_{11}$, to $E_{22}$ and to $E_{33}$ the numerical value of -1. In our scheme this is to say that we have an idempotent $\psi$ such that

$$(E_{11}+1)\psi = 0 \;;\; (E_{22}+1)\psi = 0 \;;\; (E_{33}+1)\psi = 0 \tag{3.18}$$

The idempotent $\psi$ is $\psi_1\psi_2\psi_3$ with

$$\psi_1 = \frac{(E_{11}-1)}{2} \;;\; \psi_2 = \frac{(E_{22}-1)}{2} \;;\; \psi_3 = \frac{(E_{33}-1)}{2} \tag{3.19}$$

and

$$E_{11}\psi = -\psi \;;\; E_{22}\psi = -\psi \;;\; E_{33}\psi = -\psi \tag{3.20}$$

In addition we have that

$$\psi_1\psi_2\psi_3 = \psi_2\psi_1\psi_3 = \psi_3\psi_1\psi_2 \tag{3.21}$$

Let us look again to the (2.77). As previously said, they represent all the algebraic sets that we may form when we consider basic elements at n=4.

Starting with the (3.20) that, as we outline again, represents the core of the EPR argument, we start to calculate:

$$E_{0i}(E_{11}+1)\psi = 0 \;;\; E_{i0}(E_{11}+1)\psi = 0 \;;$$

$$E_{ii}(E_{11}+1)\psi = 0 \;;\; E_{ij}(E_{11}+1)\psi = 0 \tag{3.22}$$

We will calculate also similar expressions for $E_{22}$ and still for $E_{33}$.

After such calculations we obtain some important results. They are given in the following manner:

$$E_{01}\psi = -E_{10}\psi = -iE_{23}\psi \;; \tag{3.23}$$

$$E_{02}\psi = -E_{20}\psi = iE_{13}\psi \;; \tag{3.24}$$

$$E_{03}\psi = -E_{30}\psi \;;\; E_{03}\psi = iE_{21}\psi \;; \tag{3.25}$$

$$E_{12}\psi = -E_{21}\psi \;; \tag{3.26}$$

$$E_{23}\psi = -E_{32}\psi \;; \tag{3.27}$$

$$E_{13}\psi = -E_{31}\psi \;; \tag{3.28}$$

Consider the (3.23-3.28), they reassume with extraordinary mathematical and physical rigor all that happens during EPR. In addition the (3.23-3.28) add also new equations, formally and conceptually unknown in traditional quantum mechanics. They represent all the algebraic quantum like relations that must potentially hold as counter part of a quantum system of two ½ spin particles in a singlet state. Let us observe that in seventy years of research on EPR, we were substantially unable to evidence in an algebraic manner all that is holding in EPR systems.

Let us examine in detail some equations.

The conclusion that

$$E_{0i} = -E_{i0} \text{ (with respect to } \psi) \tag{3.29}$$

is in perfect accord with the physical features of the considered EPR quantum system.

We have also that always it must be

$E_{01} = -iE_{23}$ ; $E_{02} = iE_{13}$ ; $E_{03} = iE_{21}$ (3.30)

and

$E_{13} = -E_{31}$ and $E_{23} = -E_{32}$ (3.31)

Finally, we have also the most relevant equation for our work. It must be that

$E_{12} = -E_{21}$ (3.32)

Its particular relevance does not arise from particular notations of algebraic or physical importance but from the fact that Asher Peres, ( first by the (3.14 and the (3.15)), arrived to formulate the EPR paradox or the algebraic contradiction as he called it correctly. We have to explain the (3.32) if we aim to give explanation of the paradox.

Let us consider the basic elements $E_{12}$ and $E_{21}$. From the algebraic context that we have introduced until here, it is evident that we may express such two basic elements in different algebraic manners. Among such different representations, the most simple, clear, but also responding (as its counter part) to a too ingenuous realistic vision, is to write such two elements in their basic forms that are:

$E_{12} = E_{10}E_{02}$ and $E_{21} = E_{20}E_{01}$ (3.33)

In this case we are ingenuously admitting as existing " elements of reality" but we are ignoring the complexity of such reality. In fact we observe immediately that the (3.33) goes in contradiction since, observing the (3.29), we have that $E_{01} = -E_{10}$ and $E_{02} = -E_{20}$

In brief , we have that

$E_{12} = E_{10}E_{02} = E_{01}E_{20} = E_{21}$ (3.34)

and

$E_{21} = E_{20}E_{01} = E_{02}E_{10} = E_{12}$

that is in contradiction with the (3.32) that instead is rigorously required in our algebraic formulation and in EPR. So we arrive to the conclusion and thus to the explanation of the paradox via our algebraic quantum like formulation.

We are engaged in an algebraic calculation that peremptorily requires to respect its basic features of non commutativity and link among sets. In other terms, in order to obtain that

$E_{12} = -E_{21}$ (3.35)

we cannot ignore that they must necessarily contain almost an explicit permutation of (1,2,3). This is to say that peremptorily we must account for non commutativity.

Making clear the permutations, this time we obtain

$$E_{12} = \frac{E_{13}E_{01}}{i} \quad (3.36)$$

Performing the same operation we also obtain that

$$E_{21} = \frac{E_{22}E_{03}}{i} \quad (3.37)$$

In this manner, $E_{12}$ may be written in the following terms

$$E_{12} = \frac{E_{10}E_{03}E_{01}}{i} = E_{10}\frac{E_{03}E_{01}}{i} \quad \text{and} \quad E_{02} = \frac{E_{03}E_{01}}{i} \tag{3.38}$$

while $E_{21}$ will be written in the following manner

$$E_{21} = \frac{E_{20}E_{02}E_{03}}{i} = E_{20}\frac{E_{02}E_{03}}{i} \quad \text{and} \quad E_{01} = \frac{E_{02}E_{03}}{i} \tag{3.39}$$

We may now solve the (3.38) and the (3.39). We find that

$$E_{12} = \frac{E_{10}E_{03}E_{01}}{i} = \frac{E_{03}E_{10}E_{01}}{i} = -\frac{E_{03}}{i} \tag{3.40}$$

and

$$E_{21} = \frac{E_{22}E_{03}}{i} = \frac{E_{20}E_{02}E_{03}}{i} = \frac{-E_{03}E_{02}E_{20}}{i} = \frac{E_{03}}{i} \tag{3.41}$$

that is to say that

$$E_{12} = -E_{21} \tag{3.42}$$

as required in EPR.

In conclusion the EPR paradox is explained in our quantum like algebraic version.

*We may conclude outlining the particular importance to have expressed* $E_{02}$ *and* $E_{01}$ *as in (3.38) and in (3.39).*

*The expression*

$$E_{02} = \frac{E_{03}E_{01}}{i} \tag{3.43}$$

*in* $E_{12}$ *given in (3.34) outlines that , if we consider* $E_{10}$ *and* $E_{12}$*, we cannot escape to consider the functional dependence (given by permutation and thus non commutativity) of* $E_{02}$ *from* $E_{01}$ *and* $E_{03}$*. In the same manner, the presence of*

$$E_{01} = \frac{E_{02}E_{03}}{i} \tag{3.44}$$

*in* $E_{21} = E_{20}E_{01}$ (3.45)

*states that, considering* $E_{20}$ *and* $E_{21}$*, we cannot escape to consider the functional dependence of* $E_{01}$ *from* $E_{02}$ *and* $E_{03}$.

The algebraic features of EPR paradox are now explained.

## 3.2.b On Some Features of Bell inequality discussed by using Clifford algebra.

There are well known the excellent expositions that were given by T.F Jordan [3] and Mermin [17] on Bell inequalities. We will follow these authors but using Clifford algebra.

For the spin of one particle, we use $E_{0i}$ ($i$ = 1, 2, 3); for the spin of the other particle, we use $E_{i0}$ ($i$ = 1, 2, 3). Owing their commuativity, we are sure that they can be measured together.

Let us admit now that we want to discuss experiments done with two particles in a state where the total spin is zero. We assume that the total spin is the vector quantity whose projections in the three perpendicular reference directions. As algebraic counterpart we use the following Clifford members.

$$\frac{1}{2}\hbar(E_{01}+E_{10}); \qquad \frac{1}{2}\hbar(E_{02}+E_{20}); \qquad \frac{1}{2}\hbar(E_{03}+E_{30}); \tag{3.46}$$

For any real numbers $x_1, x_2, x_3$, the quantity represented by the Clifford member

$$x_1(E_{01}+E_{10}) + x_2(E_{02}+E_{20}) + x_3(E_{03}+E_{30}) \tag{3.47}$$

has the value zero. The projection of the total spin in any direction is zero. Thus, we have

$$<(E_{01}+E_{10})^2>=0; \quad <(E_{02}+E_{20})^2>=0; <(E_{03}+E_{30})^2>=0$$

and $(E_{01}+E_{10})^2 = E_{01}^2 + E_{10}^2 + 2E_{01}E_{10}$ (3.48)

Consequently , we have that (3.49)

$$< E_{01} E_{10} > = -1 \tag{3.50}$$

and similarly

$$< E_{02} E_{20} > = -1 \quad \text{and} \quad < E_{03} E_{30} > = -1 \tag{3.51}$$

We have also

$$< E_{02} E_{10} > =< E_{01} E_{20} >=< E_{02} E_{30} >=< E_{03} E_{20} > =$$
$$< E_{03} E_{10} > = < E_{01} E_{30} > = 0 \tag{3.52}$$

Consider now two Clifford members $U$ and $V$ with $U^2 = 1$ and $V^2 = 1$, and let us admit still that

$$U V + V U = 0 \tag{3.53}$$

For any state we have that

$$< U >^2 + < V >^2 \leq 1 \tag{3.54}$$

For the proof see ref. [3,8]

Let us consider the component ($E_{01}$ + $E_{10}$) given in the (3.46) for a single state. Let us examine the (3.53-3.54) with

$$U = E_{01} E_{10} \quad \text{and} \quad V = E_{02} E_{10} \tag{3.55}$$

Let us verify that we have

$$U^2 = V^2 = 1 \ ; \quad U V + V U = 0 \tag{3.56}$$

as required. Let us now apply the (3.50), and the (3.51), it is

$$< E_{02} E_{10} > = 0 \tag{3.57}$$

Still, consider

$$U = E_{01} E_{10} \quad \text{and} \quad V = E_{03} E_{10} \tag{3.58}$$

we have

$$< E_{03} E_{10} > = 0 \tag{3.59}$$

Let us repeat our procedure for ($E_{02}$ + $E_{20}$). We have

$$< E_{02} E_{20} > = -1 \tag{3.60}$$

Let us consider

$U = E_{02}E_{20}$ and $V = E_{01}E_{20}$ (3.61)

Again, we have that

$U^2 = V^2 = 1$ and $UV + VU = 0$ (3.62)

and we obtain that

$< E_{01}E_{20} > = 0$ (3.63)

Let us consider now the following Clifford Members

$A = E_{02}E_{10}$ and $B = E_{01}E_{20}$ (3.64)

Let us write $AB = E_{02}E_{01}E_{10}E_{20} = E_{03}E_{30} = E_{30}E_{03}$. (3.65)

According to the (3.51), we have that

$< AB > = < E_{03}E_{30} > = -1$ (3.66)

Consider now the following Clifford members

$E_{0i} = E_{i0}E_{ii}$ or $E_{i0} = E_{0i}E_{ii}$ ; $i = 1$ or $2$ or $3$ (3.67)

We see that the Clifford basic element $E_{ii}$ is here what we could call the specific "non local" Clifford algebraic coupling for the spin system ($E_{0i}$ , $E_{i0}$) taken in consideration.

To complete our exposition, let us return briefly to consider the following Clifford expression

$E_{01}E_{10}E_{02}E_{10} + E_{02}E_{10}E_{01}E_{10}$ (3.68)

It is immediate to verify that we have

$E_{01}E_{10}E_{02}E_{10} + E_{02}E_{10}E_{01}E_{10} = E_{01}E_{02}E_{10}E_{10} + E_{02}E_{01}E_{10}E_{10} = 0$ (3.69)

Therefore, let

$U = E_{01}E_{10}$ ; $V = E_{02}E_{10}$ (3.70)

we have that

$U^2 = (E_{01}E_{10})^2 = 1$ ; $V^2 = (E_{02}E_{10})^2$ (3.71)

and

$< E_{01}E_{10} >^2 + < E_{02}E_{10} >^2 \leq 1$ (3.72)

For spin zero, we have that

$< E_{01}E_{10} > = -1$, and $< E_{02}E_{10} > = 0$ (3.73)

and similarly for the other results in (3.52)

Note that we have also

$< E_{01} > = < E_{02} > = < E_{03} > = 0$ ; $< E_{10} > = < E_{20} > = < E_{30} > = 0$ (3.74)

Let $a_1, a_2, a_3$ be real numbers such that

$a_1^2 + a_2^2 + a_3^2 = 1$ (3.75)

We have the Clifford member

$A(a) = a_1 E_{01} + a_2 E_{02} + a_3 E_{03}$ (3.76)

Measuring the projection of the magnetic moment in that direction for the particle determines a value, either $+1$ or $-1$, for the quantity represented by the Clifford member

$A(a) = a_1 E_{10} + a_2 E_{20} + a_3 E_{30}$ (3.77)

we have that

$< a_1E_{01} + a_2E_{02} + a_3E_{03} > = a_1 < E_{01} > + a_2 < E_{02} > + a_3 < E_{03} > = 0$ (3.78)

with probabilities

$$p(1) = \frac{1}{2} + \frac{1}{2}( < a_1E_{01} + a_2E_{02} + a_3E_{03} >) \quad (3.79)$$

$$p(-1) = \frac{1}{2} - \frac{1}{2}( < a_1E_{01} + a_2E_{02} + a_3E_{03} >) \quad (3.80)$$

We have also that

$$< a_1E_{10} + a_2E_{20} + a_3E_{30} > = a_1 <E_{10} > + a_2 <E_{20} > + a_3 <E_{30} > = 0 \quad (3.81)$$

so that there are equal probabilities ½ for the two possible values $+1$ and $-1$ for the quantity represented by the (3.77)

Now, let us assume that we decide to measure projections of the two magnetic moments in different directions. Let $b_1, b_2, b_3$ be real numbers such that

$$b_1^2 + b_2^2 + b_3^2 = 1 \quad (3.81a)$$

Measuring the projection of the magnetic moment of one particle in the direction of the $a_1, a_2, a_3$ vector and the projection of the magnetic moment of the other particle in the direction of the $b_1, b_2, b_3$ vector determines a value, either $+1$ or $-1$, for each of the two quantities represented by the Clifford members

$$A(a) = a_1E_{01} + a_2E_{02} + a_3E_{03}$$

$$(3.82)$$

$$B(b) = b_1E_{10} + b_2E_{20} + b_3E_{30}$$

Both $A(a)$ and $B(b)$ are the well known dichotomic observables used in the usual elaboration of quantum mechanics. There are non zero probabilities for all four possible pairs of values. In order to have detailed information on the valuable values, we need to consider the product of the two quantities. We need to calculate

$$< A(a)\, B(b) > \quad (3.83)$$

Calculating $A(a)\,B(b)$ we have that

$$A(a)\,B(b) = a_1b_1E_{01}E_{10} + a_2b_2E_{02}E_{20} + a_3b_3E_{03}E_{30} + a_1b_2E_{01}E_{20} + a_1b_3E_{01}E_{30} + $$
$$+ a_2b_1E_{02}E_{10} + a_2b_3E_{02}E_{30} + a_3b_1E_{03}E_{10} + a_3b_2E_{03}E_{20} \quad (3.84)$$

Its mean value is

$$< A(a)\,B(b) > = a_1 b_1 <E_{01} E_{10} > + a_2 b_2 <E_{02} E_{20}> + a_3 b_3 <E_{03} E_{30}> +$$
$$+ a_1 b_2 <E_{01} E_{20}> + a_1 b_3 <E_{01} E_{30}> + a_2 b_1 <E_{02} E_{10}> + a_2 b_3 <E_{02} E_{30} > +$$
$$+ a_3 b_1 <E_{03} E_{10} > + a_3 b_2 <E_{03} E_{20} > \quad (3.85)$$

Using the (3.52), we obtain the final value

$$< A(a)\,B(b) > = -a_1 b_1 - a_2 b_2 - a_3 b_3 \quad (3.86)$$

as it is required for standard quantum mechanics.

Finally, let us observe that $A(a)$ $B(b)$ may assume the values *1* or *– 1*. Indicated by $p(1)$ the probability to have the value 1 and by $p(-1)$ the probability to have the value - *1*, we have that

$$p(1) = \frac{1 + a_1b_1 + a_2b_2 + a_3b_3}{2}$$

$$p(-1) = \frac{1 - a_1b_1 - a_2b_2 - a_3b_3}{2}$$

In this manner we have now completed our algebraic treatment, having of course a detailed physical counterpart.

Obviously we may write that

$$p(1,1) = \frac{1}{2}p(1);\ p(-1,-1) = \frac{1}{2}p(1);\ p(1,-1) = \frac{1}{2}p(-1);\ p(-1,1) = \frac{1}{2}p(-1);$$

$$p(1) = p(1,1) + p(-1,-1);\ p(-1) = p(-1,1) + p(1,-1);$$

$$p(1,1) + p(1,-1) = \frac{1}{2};\ p(-1,1) + p(-1,-1) = \frac{1}{2};$$

$$p(1,1) + p(-1,1) = \frac{1}{2};\ p(1,-1) + p(-1,-1) = \frac{1}{2}$$

and we may examine all the experimental situations reproducing Bell inequalities.

## 4. On Some Cognitive Features of Quantum Mechanics Deduced by Using Clifford Algebra.

We will attempt now to assemble the basic features that we have previously developed**.** Before to proceed in this direction, we would add still some other comment.

First of all we have never mentioned the manner in which the time evolution of quantum system may receive consideration in a bare bone skeleton of quantum mechanics elaborated by using the Clifford algebra.

The problem to write a Schrödinger equation by using Clifford algebra was developed by us previously [8] as well as that one to write the Dirac equation.

As we know, in deriving his equation Dirac had the problem to extract the square root of

$$E^2 = p^2c^2 + m^2c^4 \tag{4.1}$$

We know that Dirac solved the problem by introducing standard well known matrices that of course, looking at the (2.27), result to be the following Clifford algebraic elements:

$$E_{11}=\begin{pmatrix}0&0&0&1\\0&0&1&0\\0&1&0&0\\1&0&0&0\end{pmatrix}\quad;\quad E_{12}=\begin{pmatrix}0&0&0&-i\\0&0&i&0\\0&-i&0&0\\i&0&0&0\end{pmatrix};E_{13}=\begin{pmatrix}0&0&1&0\\0&0&0&-1\\1&0&0&0\\0&-1&0&0\end{pmatrix}\quad;$$

$$;E_{30}=\begin{pmatrix}1&0&0&0\\0&1&0&0\\0&0&-1&0\\0&0&0&-1\end{pmatrix};\tag{4.2}$$

Generally speaking , we may do use also of the Linear Transformations that we introduced in the (1.14a). Consider the Clifford algebraic members

$$U = q_0 - i\,k\,T \text{ with } U^+ = q_0 + i\,k\,T \tag{4.3}$$

where $q_0$ and $T$ are Clifford members and $k$ is a small parameter ($k\to 0$) representing the time.

We have that

$$U\,U^+ = U^+U = q_0^2 + i\,k\,(q_0\,T - T\,q_0) + k^2T^2 = 1 \tag{4.4}$$

under the condition that $q_0^2 = 1$ ; $q_0\,T = T\,q_0$ and the limit $k\to 0$.

As a particular case we may consider $q_0 \equiv 1$, and we have that

$$U = 1 - i\,k\,T\ ;\ U^+ = 1 + i\,k\,T \tag{4.5}$$

Let us consider that we have a Clifford member $B$ representing a given physical quantity at the initial time. The time evolution of $B$, representing this quantity at a time $k$ later, is given by a Linear Transformation in the following manner

$$B_k = U\,B\,U^+ = B + i\,k\,(BT - TB) = B + i\,k\,[B, T] \tag{4.6}$$

If $B$ commutes with $T$, we have that

$$U\,B\,U^+ = B\,U\,U^+ = B \tag{4.7}$$

that is to say that $B$ is constant in time.

Let us consider the Clifford members $P_i$ ($i = 1, 2, 3$), representing the momentum, in the case of an isolated quantum system in which the momentum must be constant in time. We have that

$$\begin{aligned}(1 - i\,k\,T)\,P_1\,(1 + i\,k\,T) &= P_1\\ (1 - i\,k\,T)\,P_2\,(1 + i\,k\,T) &= P_2\\ (1 - i\,k\,T)\,P_3\,(1 + i\,k\,T) &= P_3\end{aligned}\tag{4.8}$$

The position coordinates are expressed instead by the Clifford members $Q_i$ ($i = 1, 2, 3$) at some time $k$. During a small time interval $k$ , $Q_i$ must change as

$$Q_1 \to Q_1 + \frac{1}{m}P_1\,k\,,\quad Q_2 \to Q_2 + \frac{1}{m}P_2\,k\,,\quad Q_3 \to Q_3 + \frac{1}{m}P_3\,k\,, \tag{4.9}$$

We have that

$$(1 \mp i\,k\,T)\,Q_1\,(1 \pm i\,k\,T) = Q_1 \mp \frac{1}{m}P_1\,k$$

$$(1 \mp i\,k\,T)\,Q_2(1 \pm i\,k\,T) = Q_2 \mp \frac{1}{m} P_2\,k \tag{4.10}$$

$$(1 \mp i\,k\,T)\,Q_3(1 \pm i\,k\,T) = Q_3 \mp \frac{1}{m} P_3\,k$$

We have that it must be $[P_i, T] = 0$
and

$$[Q_i, T] = +\,i\frac{p_i}{m} \tag{4.11}$$

with

$$Q_i\,P_i - P_i\,Q_i = i\,\hbar\,e_3\text{, with } e_3 = +1 \text{ in } N_{i,+1} \tag{4.12}$$

The previous equations agrees with standard quantum mechanics, if we consider $T$ to be the following Clifford Member

$$T = \frac{1}{\hbar} H \tag{4.13}$$

With

$$H = \frac{1}{2m}(P_1^2 + P_2^2 + P_3^2) \tag{4.14}$$

that is the Clifford member representing the hamiltonian of the system, As a general rule, we may introduce two Clifford members

$$a = [V(Q)]^{½} + i\,\frac{P}{(2m)^{1/2}} \qquad a^{+} = [V(Q)]^{½} - i\,\frac{P}{(2m)^{1/2}} \tag{4.15}$$

in which $V(Q)$ represents the potentials.
The hamiltonian of the system is given by the following manner

$$H = \frac{1}{2}(a\,a^{+} + a^{+}\,a) = \frac{P^2}{2m} + V(Q) \tag{4.16}$$

In conclusion, we may reconstruct the bare bone skeleton relating spin rotations, changes in space location, changes in momentum and, finally, all the well established chapter of the invariances in standard quantum mechanics.
In conclusion, we have given a rather complete and satisfactory bare bone skeleton of quantum mechanics by using the Clifford algebra.
With the only exception of the derivations going from the (4.3) to the (4.16) the very surprising feature of our exposition is that we arrived to give a quite complete algebraic exposition of quantum mechanics without invoking never physical principles.
Obviously we were not able to introduce a justification of quantization and thus of $\hbar$.
Here we used only the basic rules of Clifford algebra, in particular the $A(Si)$ algebra as well as the $N_{i,\pm 1}$. We did not use other principles, and, in particular, we did not utilize results of physics.

It is important to outline that we obtained the manner to describe changes in time, space, and in momentum, and, finally, (what it is of basic importance for us), we had also the relative invariances.
We have to give a detailed interpretation of this basic feature, and to this purpose we have to recall what we analyzed previously.
We must ask what is the reason to attribute a so large importance to idempotents. In our elaboration. The answer is in perfect accord with von Neumann basic statement. He showed that projection operators (that from this moment we will indicate by $\Lambda$) (that are the idempotents of Clifford algebra) represent logical statements. Consequently, in standard quantum physics as well as in our bare bone skeleton of quantum mechanics we have the constant presence of idempotents:

a) such idempotents are members of Clifford algebra;
b) Idempotents represent logical statements. By obvious extension, we say that they represent conceptual entities.

To reinforce such our view point we must remember here that in standard quantum mechanics, given an observable $A$ admitting possible eigenvalues $k_1, k_2,........$ we have

$$A = \sum_{k_i} k_i \Lambda_{k_i} \tag{4.17}$$

Certainly, all we well know the von Neumann quantum mechanical mathematical foundations, spectral theorem, its meaning and its profound link with standard axiomatic formulation in Hilbert space. However, the particular question that we pose here is rather different. We demand how is that in the (4.17) we find as profound linked the values that the observable $A$ may assume (object) and logical statements (thus, conceptual entities) as represented by the $\Lambda_{k_i}$. To this purpose we outlined previously that in quantum mechanics if $|\psi\rangle$ is a state vector for a quantum state in which the observable $S$ is equal to $k$ ,than the density matrix $|\psi\rangle\langle\psi|$ represents the logical statement $\Lambda_k$ which says "$S = k$".
Consequently we insist demanding what is the reason of such profound linking in quantum mechanics (as in particular evidenced by our Clifford elaboration) between "physical features" and logical statements that of course mean cognition and thus cognitive performance, and, in particular, presence of conceptual entities.
In brief: we have derived quantum mechanics using only the Clifford algebra and its few basic axioms and principles. As seen, basic Clifford algebraic principles relate logical statements and thus the logic and the cognition. We have to clear in detail the reason because quantum mechanics involves directly conceptual entities.
We retain that a very distinguished scientist has formulated in detail this question in the years of his application It is absolutely necessary and indispensable to fasten his ideas and his basic results in the present our exposition of quantum mechanics elaborated by the Clifford algebra. To this purpose, one of his fundamental paper entitles " The logical Origins of Quantum mechanics". Just the title of the paper clearly indicates the way we must pursue in order to actually understand the foundations of quantum theory.
The object of our interest is the following question.
We may say that the basic features of quantum mechanics result to be consistent with the logic formulation that we introduced by using the Clifford algebra. We repeat here, J. von Neumann showed that projection operators of quantum mechanics can be interpreted as logical statements. We may say that he constructed a matrix logic derived from quantum mechanics. As we will show soon after, we may reach instead an inverted objective. By using Clifford algebra, we may show that quantum mechanics is constructed on the basis of a logic realized by Clifford algebra.
In other terms, the situation results to be inverted, not logic from quantum mechanics but quantum mechanics from logic. Consequently, it arises as necessary the problem to ask what are the reasons because in quantum mechanics the logic statements become "observables" themselves. Still, following this line of exposition, it derives that we have a logical relativism in quantum mechanics while instead such relativism does not exist in classical physics. We have to explain also such important feature. Finally, we have to clear the reason because logic, and thus language, semantic and human cognition, play such a fundamental role in quantum mechanics while only an auxiliary support may be found in classical physics. Still according in some manner with Orlov [7], we may explain such features considering the fundamental conclusion of the present paper:
***There are stages of our reality in which we no more can separate the logic (and thus cognition and thus conceptual entity) from the features of "matter per se". In quantum mechanics the logic, and thus the cognition and thus the conceptual entity-cognitivism, assumes the same importance as the features of what is being described. We are at levels of our reality in which the truths of logical statements about dynamic variables become dynamic variables themselves so that a profound link is established from its starting in this theory between physics and logic***.
***Finally, in this approach there is not an absolute definition of logical truths. Transformations , and thus … "redefinitions" of truth values are permitted in such scheme as well as the (1.14a) as well as the previously identified invariance principles, clearly indicate.***
This is the basic conclusion that we reach in this paper. There are several results that have been included , and we aim to examine them step by step.
Y. Orlov gave an excellent example of this fundamental features of quantum reality observing that in quantum mechanics we have observables linked to dynamic physical variables (call one by Q, as example.). On the other hand, we have the corresponding projectors $\Lambda_Q$ (elements of the Clifford algebra)that of course represent logical statements. The $\Lambda_Q$ always commutes with $\hat{Q}$ that is the hermitean operator connected to $Q$.
Let us consider now that the algebra $A(S_i)$ admits idempotents.

Two of such idempotents, as repeatedly mentioned, are, as example,

$$\psi_1 = \frac{1+e_3}{2} \quad \text{and} \quad \psi_2 = \frac{1-e_3}{2} \qquad (4.18)$$

A generic member of algebra $A(S_i)$ is given by

$$x = \sum_{i=0}^{4} x_i e_i$$

with $x_i$ pertaining to some field $\Re$ or $C$.

We may transform Clifford members of $A(S_i)$ using linear homogeneous transformations so that

$$x' = SxS^+ \qquad (4.18a)$$

Generally speaking we may take

$$S(a,b) = \frac{a+a^*}{2} + (\frac{b+b^*}{2})e_1 + (\frac{i(b-b^*)}{2})e_2 + (\frac{a-a^*}{2})e_3 \qquad (4.19)$$

$$S^+(a,b) = \frac{a+a^*}{2} + (\frac{b^*-b}{2})e_1 - (\frac{i(b+b^*)}{2})e_2 + (\frac{a^*-a}{2})e_3 \qquad (4.20)$$

(13)

that are members of the $A(S_i)$ algebra. Taking the relative matrix representation, one acknowledges easily that we are considering the $SU_2$ group.

In analysis of Clifford algebraic $A(S_i)$ transformed members we usually use [8] the following transformation

$$S(a,b) = a + b\,e_i \quad ; \quad S^+(a,b) = a^* + b^* e_i \quad ; \quad i=1,2,3 \qquad (4.21)$$

with

$$aa^* + bb^* = 1 \quad \text{and} \quad ab^* + a^*b = 0 \qquad (4.22)$$

and

$$SS^+ = S^+S = 1$$

Applying the (4.21) we obtain the three new Clifford basic elements:

$$\bar{e}_1 = S\,e_1\,S^+ \quad ; \quad \bar{e}_2 = S\,e_2\,S^+ ; \quad \bar{e}_3 = S\,e_3 S^+ \qquad (4.23)$$

that in matrix notation assume the following forms

$$\bar{e}_1 = \begin{pmatrix} 0 & \omega \\ \vartheta & 0 \end{pmatrix} , \quad \bar{e}_2 = \begin{pmatrix} 0 & -i\omega \\ i\vartheta & 0 \end{pmatrix} , \quad \bar{e}_3 = \begin{pmatrix} 1 & 0 \\ 0 & -1 \end{pmatrix} = e_3 \qquad (4.24)$$

with

$$aa^* - bb^* - 2a^*b = \vartheta \quad \text{and} \quad aa^* - bb^* + 2a^*b = \omega \qquad (4.25)$$

Obviously, the (4.23) or the (4.24) represent a new basic triad of Clifford algebraic elements in $A(S_i)$.

The first, elementary (atomic) statement $\lambda_k$ is represented it by $2\times 2$ diagonal matrix

$$\lambda_k = \begin{pmatrix} 1 & 0 \\ 0 & 0 \end{pmatrix} \quad , \quad \lambda_k^2 = \lambda_k \tag{4.26}$$

$\lambda_k$ is a statement that may be true (eigenvalue +1) or false (eigenvalue 0 ) and the negation of $\lambda_k$ is $\bar{\lambda}_k$ represented by the matrix

$$\bar{\lambda}_k = 1 - \lambda_k = \begin{pmatrix} 0 & 0 \\ 0 & 1 \end{pmatrix} \quad ; \quad \bar{\lambda}_k^2 = \lambda_k \tag{4.27}$$

Consider:

$$\lambda_0 = \frac{1+e_3}{2} \tag{4.28}$$

that in our logic scheme by Clifford $A(S_i)$ algebra, represents the elementary logic statement. Other idempotents, and thus other logical statements, are given by supporting the following conditions

$$a = 1/2 \quad , \quad b = 0 \quad , \quad c = \pm\frac{sen\gamma}{2} \quad , \quad d = \pm\frac{\cos\gamma}{2}$$

or

$$a = 1/2 \quad , \quad b = \pm\frac{sen\gamma}{2} \quad , \quad c = 0 \quad , \quad d = \pm\frac{\cos\gamma}{2} \tag{4.29}$$

or still

$$a = 1/2 \quad , \quad b = \pm\frac{sen\gamma}{2} \quad , \quad c = \pm\frac{\cos\gamma}{2} \quad , \quad d = 0$$

and

$$a = 1/2 \quad , \quad b = c = d = \frac{1}{2\sqrt{3}}$$

Looking at the (4.29), we deduce that, generally speaking, we may construct several, different and more complex statements

$\lambda_0, \lambda_1, \lambda_2, \ldots\ldots\ldots\ldots\ldots\ldots, \lambda_n$

that may result to be true or false or in their potential state of logical indetermination until a numerical value is attributed directly to them by the theorem 2. By subsequent application of the (4.18), we will have

$$\lambda_0 \to \lambda_j \to \cdots\cdots\lambda_k \tag{4.30}$$

and, for each pair of logical statements, we may calculate $\lambda_j\lambda_0$ that will be still an element of the $A(S_i)$ Clifford algebra. Applying to such member the theorem n.2a or n.2b in $N_{i,\pm 1}$, we will calculate the probability that the logical statement $\lambda_j$ may be predicted to be true or false starting with the truth value of $\lambda_0$. Call the logical statement $\lambda_0$ by $A$, and we mean that it is true when we write $A = +1$. Otherwise, we intend that it is false, when we write $A = -1$. Call the logical statement $\lambda_j$ by $B$, being it true when we write $B=+1$, and false when $B=-1$. By the procedure of application of theorem n.2a, n.2b to $\lambda_j \lambda_0$, and thus passing from $A(S_i)$ to $N_{i,\pm 1}$, assuming as example $A$ to be true, we will calculate the probability that the logical statement $B$ is predicted to be true being $A$ true, or the probability that the logical statement $B$ is predicted to be false being $A$ true. This is to say that by this procedure we evaluate the probabilities $p(B=+1/A=+1)$ or $p(B=-1/A=+1)$.

Note two important features that we are delineating by our Clifford logic scheme. Let us admit that we estimate $\lambda_1 \lambda_0$ as member of $A(S_i)$. Passing from $A(S_i)$ to $N_{i,\pm 1}$, two possible cases will result possible. Or we will obtain a new element in $N_{i,\pm 1}$ that will be reduced directly to a numerical value, and consequently we will conclude to have directly estimated the probability $p(B/A)$ or it will result instead that we obtain an element of $N_{i,\pm 1}$ that of course does not assume directly a numerical value.

**This is a case in which we will conclude for what we will call the *incompetence* of the logical statement $\lambda_0$ to predict probability for true or false value of $\lambda_1$.**

As we will see in detail in the subsequent calculations, we will estimate Clifford members as $\lambda_1 \lambda_0$, $\lambda_2 \lambda_1$, computing each time the classical probabilities $p(B/A)$, $p(C/B)$ and obviously we will expect at the end to find the same calculated value of probability, $p(C/A)$, corresponding to $\lambda_2 \lambda_0$. In performing calculations corresponding to such classical scheme of probability, each time (that is to say in $\lambda_1 \lambda_0$, and $\lambda_2 \lambda_1$), we will force the subordinate logical statement to result or true or false. Following this procedure , we will exclude a priori the important cases in which the *incompetence* for the subordinate statement to be predicted might appear. Computing instead directly $\lambda_2 \lambda_0$, it will remain free the possibility for the statement $\lambda_0$ to have *incompetence* in predicting probability for $\lambda_1$

We may now pass to the calculations.

Let us start considering the following basic statement

$$\lambda_0 = \frac{1+e_3}{2} \tag{4.31}$$

and let us calculate the new logical statement $\lambda_1$

$$\lambda_1 = S\,\lambda_0 S^+ \tag{4.32}$$

with $S$ and $S^+$, Clifford algebraic elements in $A(S_i)$, given in the following manner

$$S = \cos(\beta_1/2) + i\,sen(\beta_1/2)\,e_2 \quad \text{and} \quad S^+ = \cos(\beta_1/2) - i\,sen(\beta_1/2)\,e_2 \;;$$

$$S\,S^+ = 1. \tag{4.33}$$

We obtain that

$$\lambda_1 = \frac{1}{2} - \frac{1}{2} sen\beta_1 e_1 + \frac{1}{2}\cos\beta_1\, e_3 \qquad (4.34)$$

Note that it is one of the idempotents previously identified in (4.29) .
It is a logical statement.
Obviously it is $\lambda_1^2 = \lambda_1$. Let us observe also that we have found three new basic elements

$$\bar{e}_1 = \cos\beta_1\, e_1 + sen\beta_1\, e_3\ ,\ \bar{e}_2 = e_2,\ \bar{e}_3 = -sen\beta_1\, e_1 + \cos\beta_1\, e_3 \qquad (4.35)$$

These are three new basic elements that pertain to the $A(S_i)$ algebra, obeying the standard $A(Si)$ rules .
Note of course that the logical statement $\lambda_1$ may be rewritten as

$$\lambda_1 = \frac{1+\bar{e}_3}{2} \qquad (4.36)$$

In such new scheme with three basic elements $(\bar{e}_1, \bar{e}_2, \bar{e}_3)$, $\lambda_1$ may result to be or true or false by applying to it the $N_{i,\pm 1}$
Now, obeying to the same principle, let us calculate

$$\lambda_2 = S\,\lambda_1\,S^+ \qquad (4.37)$$

where this time we have

$S = \cos(\beta_2/2) + i\,sen(\beta_2/2)\,e_2$ and $S^+ = \cos(\beta_2/2) - i\,sen(\beta_2/2)\,e_2$ ;
(4.38)
$SS^+ = 1$
We obtain that

$$\lambda_2 = \frac{1}{2} - \frac{1}{2} sen(\beta_1 + \beta_2) + \frac{1}{2}\cos(\beta_1 + \beta_2)\,e_3 \qquad (4.39)$$

Again it is one of the idempotents given in (4.29)

$$\lambda_2^2 = \lambda_2 \qquad (4.40)$$

According to our approach it is a new logical statement.
Here we have found still three new basic elements in $A(S_i)$ algebra

$$\hat{e}_1 = \cos(\beta_1 + \beta_2)e_1 + sen(\beta_1 + \beta_2)e_3\ ; \qquad \hat{e}_2 = e_2;$$
$$\hat{e}_3 = -sen(\beta_1 + \beta_2)e_1 + \cos(\beta_1 + \beta_2)e_3 \qquad (4.41)$$

These are still three new basic elements of the $A(S_i)$ algebra.
Again we may write the new logical statement

$$\lambda_2 = \frac{1+\hat{e}_3}{2} \qquad (4.42)$$

that may result true or false by using $N_{1,\pm 1}$.
We have now calculated the three Clifford elements $\lambda_0, \lambda_1, \lambda_2$ . For easiness, let use the previous scheme, and let us indicate such logic statements by $A, B, C$ .

We agree to write $A=1$ when the logical statement $A$ is true, and $A=-1$ when it is false. We adopt the same convention for the logical statements $B$ and $C$, respectively.
Considering the (4.34) and the (4.31), let us calculate the Clifford member $\lambda_1\lambda_0$. After calculations we obtain that

$$\lambda_1\lambda_0 = (\frac{1}{4}+\frac{1}{4}\cos\beta_1)+\frac{1}{4}sen\beta_1(-e_1+ie_2)+(\frac{1}{4}+\frac{1}{4}\cos\beta_1) \quad (4.43)$$

Let us observe again that this is a member of $A(S_i)$. Let us admit now that we intend to calculate the probability that $\lambda_1$ is true when $\lambda_0$ is true. This is to say $p(B=+1/A=+1)$. On the basis of the two theorems shown, the algebra $A(S_i)$ no more holds, and instead the algebra $N_{1,+1}$ with its proper rules of commutation. By application of such rules in the (4.43) we have not problems in attributing a direct numerical value to (4-43) owing to the presence of the term

$$\frac{1}{4}sen\beta_1(-e_1+ie_2)$$

that directly goes to zero.
So we finally obtain

$$\lambda_1\lambda_0 = \frac{1}{2}+\frac{1}{2}\cos\beta_1 = \cos^2(\beta_1/2) \quad (4.44)$$

In conclusion, we obtain that

$$p(B=+1/A=+1)=\cos^2(\beta_1/2) \quad (4.45)$$

Note that we have not possible *incompetence* in this case. Given the logical statement $\lambda_0$,we always may estimate the probability of the logical statement $\lambda_1$ to be true or false being $\lambda_0$ true or false
Repeating the same procedure we obtain that

$$p(B=-1/A=+1)=sen^2(\beta_1/2) \quad (4.46)$$

with

$$p(B=+1/A=+1)+p(B=-1/A=+1)=1 \quad (4.47)$$

We have not indetermination (that is to say *incompetence*) in this case.
Considering now the (4.34) and the (4.39), let us calculate the Clifford algebraic element $\lambda_2\lambda_1$. We obtain that
that

$$\begin{aligned}\lambda_2\lambda_1 &= (\frac{1}{4}+\frac{1}{4}sen\beta_1 sen(\beta_1+\beta_2)+\frac{1}{4}\cos\beta_1\cos(\beta_1+\beta_2))+\\ &(-\frac{1}{4}sen\beta_1-\frac{1}{4}sen(\beta_1+\beta_2))e_1+\frac{ie_2}{4}(\cos\beta_1 sen(\beta_1+\beta_2)+\\ &\frac{ie_2}{4}(-sen\beta_1\cos(\beta_1+\beta_2))+(\frac{1}{4}\cos\beta_1\cos(\beta_1+\beta_2)e_3\end{aligned} \quad (4.48)$$

This is still a member of $A(S_i)$ Clifford algebra.

Consider now that we intend to calculate the probability that $\lambda_2$ is true when $\lambda_1$ is true. This is to say $p(C=+1/B=+1)$.

By application of $N_{i,\pm 1}$, we obtain for the following expression

$$\lambda_2\lambda_1 = (\frac{1}{4}+\frac{1}{4}sen\beta_1 sen(\beta_1+\beta_2)+\frac{1}{4}\cos\beta_1\cos(\beta_1+\beta_2))+$$

$$(-\frac{1}{4}sen\beta_1-\frac{1}{4}sen(\beta_1+\beta_2))e_1+\frac{ie_2}{4}(\cos\beta_1 sen(\beta_1+\beta_2)+\frac{ie_2}{4}(-sen\beta_1\cos(\beta_1+\beta_2))$$

$$+(\frac{1}{4}\cos\beta_1\cos(\beta_1+\beta_2) \qquad (4.49)$$

As it is immediately verified, it no more gives a direct numeric expression since, by application of the proper rules, the terms containing this time $e_1$ and $e_2$ no more disappear. This is a case of *incompetence* for $\lambda_1$ to estimate probability for $\lambda_2$. The only way remaining to obtain a direct numerical value it is that we consider

$$sen\beta_1 = 0\ , \qquad (4.49a)$$

and in this case we have that

$$\lambda_2\lambda_1 = (\frac{1}{4}+\frac{1}{4}\cos\beta_2)+\frac{1}{4}sen\beta_2(-e_1+ie_2)+(\frac{1}{4}+\frac{1}{4}\cos\beta_2)e_3 \qquad (4.50)$$

It is still a member of $A(S_i)$. In order to calculate $p(C=+1/B=+1)$, we remember the $A(S_i)$ algebra no more holds, and instead we must use the $N_{i,+1}$. Consequently, we obtain

$$\lambda_2\lambda_1 = \frac{1}{2}+\frac{1}{2}\cos\beta_2 \qquad (4.51)$$

that is to say that

$$p(C=+1/B=+1)=\cos^2(\beta_2/2) \qquad (4.52)$$

and the probability is now estimated..

Using the same procedure for calculations, we may obtain the probability that $\lambda_2$ is true if $\lambda_1$ is false. We obtain

$$p(C=+1/B=-1)=sen^2(\beta_2/2) \qquad (4.53)$$

In conclusion we have:

$$p(B=+1/A=+1)\ p(C=+1/B=+1)+\ p(B=-1/A=+1)\ p(C=+1/B=-1)=$$

$$=\cos^2(\beta_1/2)\cos^2(\beta_2/2)+sen^2(\beta_1/2)sen^2(\beta_2/2) \qquad (4.54)$$

It remains now to calculate $p(C=+1/A=+1)$ that is to say the probability that the logical statement $\lambda_2$ is true if $\lambda_0$ is true.

Let us calculate $\lambda_2\lambda_0$ using the (4.31), and the (4.39) and the proper rules. We obtain that

$$\lambda_2\lambda_0 = (\frac{1}{4}+\frac{1}{4}\cos(\beta_1+\beta_2)) + \frac{1}{4}sen(\beta_1+\beta_2)(-e_1+ie_2) + (\frac{1}{4}+\frac{1}{4}\cos(\beta_1+\beta_2))\ e_3 \quad (4..55)$$

This is still a member of $A(S_i)$. In order to calculate $p(C=+1/A=+1)$ , the $A(S_i)$ algebra no more holds, and we have to pass to $N_{i,+1}$. We obtain

$$\lambda_2\lambda_0 = \frac{1}{2}+\frac{1}{2}\cos(\beta_1+\beta_2) \quad (4.56)$$

that is to say that

$$p(C=+1/A=+1) = \cos^2((\beta_1+\beta_2)/2) \quad (4.57)$$

Note that also in this case we have not had problems. The application of $N_{i,\pm 1}$ has enabled us to attribute a direct numerical value to

$\lambda_2\lambda_0$ and to $p(C=+1/A=+1)$ . (4.58)

However, in calculating $\lambda_2\lambda_0$, we have transformed $\lambda_0$ in $\lambda_1$ and $\lambda_1$ in $\lambda_2$. Executing such transformations we did not impose the restrictions, given in (4.49a)), that instead we introduced in (4.49) in order to eliminate *incompetence* or, that is to say, indeterminism. On this basis we calculated the (4.52).

We are now in the condition to summarize our results.

(a)- According to von Neumann who stated that projection operators and, in particular, quantum density matrices, represent logical statements, transferring such argument at an algebraic Clifford level, we have assumed that the idempotents, given in the algebra $A(S_i)$, represent logical statements with this algebra characterized by the basic rules given by the theorems n.1and n.2 .

(b)- We have also assumed that $e_3$, the third basic element of $(e_1, e_2, e_3)$ of the given $A(S_i)$ algebra, represents an atomic proposition of classical logic as well as Y. Orlov assumed in 1982 .

(c)- According to our previous papers , we have also considered that in order to evaluate the truth or false value of a given logical statement, we have to pass from the algebra $A(S_i)$ to the algebra $N_{1,\pm 1}$ whose existence is shown by the theorems n.2a and n.2b.

(d)- Still according to Orlov, we have rejected the hypothesis that at some stages of our reality it exists a definition of absolute logical truth . Instead we have admitted a principle of relativity of logical truth values, assuming that such principle is realized by using linear homogeneous transformations of Clifford algebraic $A(S_i)$ elements. We have described in detail such basic feature starting by the (4.11).

Let us give a step on in this direction.

Substantially, according to Orlov, what we admit is that it exists $\Lambda_k \to \Lambda_k^{'}$. We call it a Clifford algebraic statement transformed in another logical statement. . Orlov call this operation a transition from one logic $K$ to another logic $K^{'}$

Obviously , recalling all the previous and repeatedly mentioned rules of our algebraic approach, we must have that the repetition of a statement gives the same statement (remember the property of the idempotents $\Lambda^2 = \Lambda$).A tautology must be transformed in a tautology. In brief $\Lambda'_k = S\Lambda S^{-1}$.

(e)- There is still a statement that follows from using theorems n.1 and n2a and n.2b in our logical approach by Clifford algebra. It is that there is not exist in our reality the possibility of always defining unconditionally a truth or its relative and subordinate probability. Let us redefine better the concept of *incompetence* that we have previously delineated. Let the two statements $A$ and $B$ be represented by $\lambda_i$ and $\lambda_j$, respectively. As said, by applying $N_{i,\pm 1}$ to $\lambda_j \lambda_i$ we may calculate $p(B/A)$.

According to standard logic and reasoning, we have only two possibilities . One is the case in which we calculate the probability that being $A$ true, also $B$ is true or being $A$ false also $B$ is false. The second case is that being $A$ true, $B$ is false or being $A$ false, $B$ is true. To such before mentioned possibilities we must add also the case in which $A$ has not the *competence* to establish the logical truth values of $B$. We have in this case a situation of intrinsic indetermination.

On the basis of such assumptions, we have calculated $\lambda_1\lambda_0$ and $\lambda_2\lambda_1$, and to such members of $A(S_i)$ algebra we have applied the theorem n.2a and n.2b obtaining the probabilities

$$p(B=+1/A=+1)=\cos^2(\beta_1/2) \quad ; \quad p(B=-1/A=+1)=sen^2(\beta_1/2)$$

and still

$$p(C=+1/B=+1)=\cos^2(\beta_2/2) \quad ; \quad p(C=+1/B=-1)=sen^2(\beta_2/2) \qquad (4.59)$$

Using the same procedure of calculation and elaboration we calculated also $\lambda_2\lambda_0$ and the probability p(C=+1/A=+1) expecting to find as in classical probability theory that

$$p(C=+1/A=+1)=\cos^2(\beta_1/2)\cos^2(\beta_2/2)+sen^2(\beta_1/2)sen^2(\beta_2/2) \qquad (4.60)$$

Instead, according to the (4.57) , we found a non classical probability result that is

$$\hat{p}(C=+1/A=+1)=\cos^2((\beta_1+\beta_2)/2) \qquad (4.61)$$

In conclusion we had that

$$p(C=+1/A=+1)\neq \hat{p}(C=+1/A=+1) \qquad (4.62)$$

and

$$\hat{p}_{non.classical}(C=+1/A=+1)=p_{classical}(C=+1/A=+1)-\frac{1}{2}sen\beta_1 sen\beta_2 \qquad (4.63)$$

The (4.63) is also in accord with the results that were obtained previously by Orlov [7].

In conclusion we find the presence of the interference term

$$-\frac{1}{2}sen\beta_1 sen\beta_2 \qquad (4.64)$$

In this manner we have reached a basic result. We have given proof of quantum interference and of indeterminism by using only logic realized in the framework of Clifford algebra and adopting only the two previously mentioned theorems n.1 and n.2a and n.2b.

It is well known that quantum mechanics runs about two basic foundations that are just the indeterminism and the quantum interference. We have obtained both such two foundations without adopting physics, and, in detail, without adopting neither one of the quantum physic principles or rules that characterize the physical basis of quantum theory. Therefore, the origins of indeterminism and of quantum interference are not in physics itself but in the logic introduced in our Clifford logic scheme. Cognition not only coexist with "matter" in quantum mechanics but in some sense the first supervises the second.

There is still another important feature that we have to outline here.

To this purpose, let us perform now the last calculation. Look again to the (4.61)

$$\widehat{p}(C=+1/A=+1)=\cos^2((\beta_1+\beta_2)/2) \qquad (4.65)$$

It follows that

$$\sqrt{\widehat{p}(C=+1/A=+1)}=\cos((\beta_1+\beta_2)/2))=\cos(\beta_1/2)\cos(\beta_2/2)- sen(\beta_1/2)\,sen(\beta_2/2) \qquad (4.66)$$

that on the basis of the (4.59) may be re-written in the following manner:

$$\sqrt{\widehat{p}(C=+1/A=+1}=\sqrt{p(B=+1/A=+1}\,\sqrt{p(C=+1/B=+1}+ \sqrt{p(B=-1/A=+1)}\,\sqrt{p(C=+1/B=-1)} \qquad (4.67)$$

that is the rule of probability amplitudes in quantum mechanics.

In conclusion, we have obtained here the formalization of the main quantum phenomena, that are the indeterminism and the interference, using only conceptual entities. Consequently, our final conclusion is that our bare bone skeleton of quantum mechanics as well as the standard quantum theory involve directly conceptual entities.

A basic problem now arises.

Have we reached experimental evidences, confirming the existence of quantum interference at the level of mental states only?

The answer to this fundamental question is positive. We have performed detailed experiments in this direction. We have not the possibility to discuss such experiments here for brevity. They have been published by us in a number of papers [11] confirming the existence of the quantum interference effect at perceptive-cognitive level in humans, and we invite the reader to look at them for a detailed exposition of the argument. We have also reached evidence of Bell inequality violation in mental states.

There still remains a feature to be discussed.

Where is that standard quantum wave functions of quantum mechanics arise in our Clifford bare bone skeleton of quantum mechanics?.

Still according to Orlov, we have to recall here again clear here the principle that we outlined previously.

We rejected the hypothesis that it exists a definition of absolute logical truth. Instead we have admitted a principle of relativity of logical truth values, assuming that such principle is realized by using linear homogeneous transformations of Clifford algebraic $A(S_i)$ elements.

Let us remember that we started to expose the section n.4 of the present paper discussing in an algebraic Clifford scheme the basic operations of standard quantum mechanics as the changes in time, in space location , and in momentum, and the consequent principles of invariance.

In the course of the present paper we have examined a kind of quantum coherence between physical properties of an object and cognition we have about it. We may say that we have as example a particle coordinate in a proper given space and simultaneously we have cognition , that is to say…….. an interpretation of it. We must have coherence between such two different features. From one hand we have as example a particle coordinate in $E$ and an interpretation of such given situation in $F$ . Some such conditions imply an existence of a symmetry. We have symmetries in logical-cognitive statements about objects instead of " objects per se".

Let us follow the argument developed by Orlov. Consider the Clifford algebraic idempotent $\Lambda_p$ that is the logical statement $\Lambda_p$: " $p = p_0$" about a particle momentum $p$ (a translational invariant in the coordinate $q$-space) . It must be by itself an invariant of the translational symmetry in the same space. Such a requirement makes sense only if $\Lambda_{p_0}$, a logical statement, is simultaneously a function of coordinates, such that it does not depend on transformations $q \to q + \delta q$ . This is possible only if we consider that $\Lambda_{p_0}$ as a function of coordinates. It does not depend on $q$ at all, or rather , as we shall consider , it depends only on differences between coordinates. So we write that $\Lambda_{p_0}(q,q') = \Lambda_{p_0}(q-q')$ . Owing to the basic properties of $\Lambda_{p_0}(q-q')$ we can consider the repeatedly outlined logical equivalence

$\Lambda_{p_0}(q-q') \wedge \Lambda_{p_0}(q-q') = \Lambda_{p_0}(q-q')$. Therefore we have the logical equivalence

$$\int \Lambda_{p_0}(q'-s)\Lambda_{p_0}(s-q)ds = \Lambda_{p_0}(q'-q) . \tag{4.68}$$

We may now solve such equation obtaining that

$$\Lambda_{p_0} = \frac{1}{A}\exp(\frac{2\pi i(q'-q)}{\lambda(p_0)}) \tag{4.69}$$

where $A = \int dq$ and the wavelength $\lambda$ depends from $p_0$. The correct eigenvectors are promptly written, and we obtain

$$\psi_{p(q)} = \exp(\frac{2\pi\, iq}{\lambda(p)}) \qquad (4.70)$$

Note the excellent result that has been obtained. It has been obtained the correct wave function of standard quantum mechanics privileging only the cognitive – logical features that we have previously examined in detail. This is the strongest support about the thesis on the logical-cognitive origins of quantum mechanics . It again supports the basic result previously introduced:

***There are stages of our reality in which we no more can separate the logic (and thus cognition and thus conceptual entity) from the features of "matter per se". In quantum mechanics the logic, and thus the cognition and thus the conceptual entity-cognitivism, assumes the same importance as the features of what is being described. We are at levels of our reality in which the truths of logical statements about dynamic variables become dynamic variables themselves so that a profound link is established from its starting in this theory between physics and logic***

In brief, following Orlov, we have obtained the correct wave function where the only missing term is obviously the Planck constant $\hbar$.

It is missing here a detailed discussion on the particular stage in which the Planck constant $\hbar$ is introduced.

In other terms, it is missing here the question to consider the problem of quantization that in fact we will examine in detail in a forthcoming paper. Of course ,in this paper we may briefly outline that, looking at the (4.70), we must have in physical terms that

$$\lambda(p) \cong \frac{c}{p}$$

Where the constant $c$ is acknowledged to be $c = \hbar$ by experiments.

Orlov previously solved the problem of quantization by using canonically conjugate variables. The reader is invited to look at this elaboration. Following our previous exposition we may anticipate here that a new method of quantization may be established using the basic assumption that invariants of translation in some space may be obviously described as logical statements represented by Clifford algebraic elements that are invariants of the same translations in the same space. The standard Heisenberg commutation rules arise as well as it is demonstrated that, as example, the momentum may be represented as in standard quantum mechanics as the operator $p = -i\hbar \frac{\delta}{\delta q}$ .

In the same manner we may consider the logical statement

$\Lambda_{q_0}$ " $q = q_0$" for coordinate position of the particle.

We may write that

$$\Lambda_{q_0} = A\delta(q' - q_0)\delta(q - q_0)$$

with $A$ constant. Also in this case we find the well known wave function (the corresponding eigenvectors) results to be $\psi_{q_i}(q) = \delta(q - q_i)$